\documentclass[aps,prd,longbibliography,reprint,twocolumn,amsmath,amssymb,amsfonts,showpacs,footnote,superscriptaddress]{revtex4-1}%reprint

\usepackage{graphicx, epsfig, amssymb}
\usepackage{amsmath, amsfonts}
\usepackage{bm}
\usepackage{enumitem}
\usepackage[usenames]{color}
\definecolor{navyblue}{rgb}{0.0, 0.0, 0.5}
\usepackage{changes}
\definechangesauthor[name={Sq}, color=orange]{Sq}
\usepackage{hyperref}
\hypersetup
	{
	        colorlinks,%
		citecolor=blue,%
		linkcolor=red,%
		urlcolor=blue,%
	}

\usepackage[caption=false]{subfig}
\usepackage[utf8]{inputenc}
\usepackage{tensor}
\usepackage{soul}
\usepackage{pict2e}
\usepackage{diagbox}
\usepackage{empheq}

\usepackage{float}
\usepackage{comment}
\usepackage{array}
\usepackage{units}
\usepackage{wasysym}
\usepackage{xcolor}
\usepackage{soul}

\DeclareMathAlphabet{\mathpzc}{OT1}{pzc}{m}{it}

\newcommand{\be}{\begin{equation}}
\newcommand{\ee}{\end{equation}}
\newcommand{\beal}{\begin{aligned}}
\newcommand{\eeal}{\end{aligned}}
\newcommand{\bea}{\begin{eqnarray}}
\newcommand{\eea}{\end{eqnarray}}

\usepackage{footnote}
\usepackage{tablefootnote}

\usepackage{threeparttable}

\begin{document}

\title{Regge pole description of scattering by dirty black holes}

\author{Theo Torres}
\email{theo.torres\_vicente@kcl.ac.uk}
\affiliation{Department of Physics, King’s College London,
The Strand, London WC2R 2LS, UK}

\author{Mohamed \surname{Ould~El~Hadj}}
\email{m.ouldelhadj@gmail.com}
\affiliation{No current affiliation, France}

\author{Shi-Qian Hu}
\email{shiqian.hu@kcl.ac.uk}
\affiliation{Department of Physics, King’s College London, The Strand,
London WC2R 2LS, UK}

\author{Ruth Gregory}
\email{ruth.gregory@kcl.ac.uk}
\affiliation{Department of Physics, King’s College London, The Strand,
London WC2R 2LS, UK}

\begin{abstract}

We study the problem of plane monochromatic scalar waves impinging upon a
Schwarzschild dirty black hole -- a Schwarzschild black hole surrounded by a thin
spherical shell of matter -- using the complex angular momentum approach.
We first recall general results concerning the differential scattering cross section
in the classical limit through a null geodesic analysis by exploring different
configurations of the shell. In particular, we show that dirty black hole spacetimes
may exhibit various critical effects for geometrical optics.  We compute the Regge
pole spectrum for various shell configurations and show that it exhibits two or three
distinct branches of poles, labelled inner surface waves, broad resonances and
outer surface waves. In the latter, two sub-families have been identified, the surface
waves associated with the outer light-ring and the creeping modes associated
with the surface of the shell. We show, using WKB analysis, that the position of the
shell sets the real part of the broad resonances while its energy-momentum and the
discontinuity of the potential at the shell’s surface set their imaginary part. Next, we
provide the complex angular momentum representation of the differential scattering
cross section and examine the role of the different Regge pole branches. We
compute the differential scattering cross section for various configurations at
several frequencies and show a very good agreement with the partial-wave
calculations. Finally, we highlight the role of the critical effects, i.e., orbiting,
glory, grazing and rainbow scattering, and their impact on the differential
scattering cross section.

\end{abstract}

\date{\today}

\maketitle

\tableofcontents

\section{Introduction}

The last decade has seen two major breakthroughs in the field of experimental astronomy.
First, the detection in 2015 of gravitational waves from a binary black hole merger
\cite{LIGOScientific:2016aoc} heralded the arrival of gravitational wave astronomy
as an experimental science. Second, the picture from the shadow of the M87
supermassive black hole from the Event Horizon Telescope
\cite{EventHorizonTelescope:2019dse} provided the first direct observation of a
black hole shadow. By providing the first direct time and spatial signals from
black holes, these measurements have taken black holes from a
purely theoretical realm into concrete physical objects in our matter-filled,
inhomogeneous Universe. Hence, it is now necessary to study black holes
not only as isolated objects but also within their environment.

This is by no means a new consideration and there is extensive literature on the
study of black hole embedded into rich and varied environments (see for example the
review articles~\cite{Barausse:2014tra,Barausse:2014pra}, motivated by the detection
of gravitational waves).
A particular emphasis was put on the impact of dark matter surrounding black holes
and their imprint of gravitational waveforms
\cite{Macedo:2013qea,Kavanagh:2020cfn,Baryakhtar:2022hbu}, (however
see also, e.g., \cite{Kiselev:2003ah,Sotiriou:2011dz,Chadburn:2013mta}
for studies of the interaction of black holes with dark energy).

Studies of black holes surrounded by matter have focused on two directions:
i) the environmental impact on the inspirals of compact objects
\cite{Macedo:2013qea,Cardoso:2019upw,Cardoso:2021wlq} and
ii) the ringdown emission~\cite{Brown:1997jv,Yoshida:2003zz,Pani:2009ss}.
The consensus on the latter was that while the resonance spectrum of black
holes surrounded with extra structures could be widely different from that of
isolated black holes, the modifications were somehow irrelevant for practical
considerations and could not be seen.
Interestingly, this conclusion is being revisited with the recent identification
and characterisation of the so-called QNMs spectral
instability~\cite{PhysRevX.11.031003,PhysRevLett.128.211102}.

In the present study, we consider the particular case of what is referred to in the
literature as a \emph{dirty black hole} (DBH), that is, a black hole surrounded by
a thin-shell of matter. DBHs were introduced as practical toy models to investigate
the impact of a local environment on black hole effects
\cite{Visser:1992qh,Visser:1993qa,Visser:1993nu}.
Surprisingly and despite the many prospects offered by gravitational wave
astronomy, the quasinormal mode (QNM) spectrum of DBHs has received
very limited attention. Generic properties of the QNM spectrum were discussed in
\cite{Medved:2003rga,Medved:2003pr} and the concrete case of a DBH was
considered by Leung et al.\ \cite{Leung:1997was,Leung:1999iq}.
In their study, Leung et al.\ adopted a perturbative approach which allowed
for an analytical investigation but restricted their conclusion to DBH spacetimes
where the shell had a limited impact.
Similar DBH configurations (that is with the shell having a limited impact on
the scattering) were investigated recently and the impact of the shell on both
the absorption and scattering cross section were computed
\cite{Macedo:2015ikq,Leite:2019uql}.

These recent studies as well as the need for concrete resonance calculations
beyond the perturbative regime are the main motivations for this study.
There are two main goals of this paper: First, to investigate wave scattering by
DBHs beyond the perturbative cases found in the literature. We therefore
consider physical DBH configurations where the shell has a substantial effect
on the scattering of waves leaving an imprint on measurable quantities.
Second, to compute the spectrum of resonances of DBHs from the Regge pole (RP)
paradigm. RPs and the associated complex angular momentum (CAM) technique
are the counterparts to QNMs. While QNMs have a real angular momentum
and a complex frequency, the RPs have a real frequency and a complex angular
momentum. This paradigm allows for a different interpretation of resonances in
a system.  The RP and CAM approaches have shed new light in many different
domains of physics involving resonant scattering theory, notably, in quantum
mechanics, electromagnetism, optics, nuclear physics, seismology and
high-energy physics (see, for example,
\cite{deAlfaro:1965zz,Newton:1982qc,Watson18,Sommerfeld49,
Nussenzveig:2006,Grandy2000,Uberall1992,AkiRichards2002,Gribov69,
Collins77,BaronePredazzi2002,DonnachieETAL2005} and references therein),
and have been successfully extended to black hole physics
\cite{Andersson_1994,Andersson_1994b,Decanini:2002ha,
Decanini:2011xi,Folacci:2018sef,Folacci:2019cmc,Folacci:2019vtt}.
As an illustration of the power of the CAM approach, we note that it provides
a unifying framework describing the glory and orbiting effects of black hole
scattering \cite{Folacci:2019cmc,Folacci:2019vtt}.

The paper is organised as follows: in Sec.\ \ref{sec:DBH_spacetime}, we review
the derivation of the DBH spacetime and show that there exist static configurations
fully characterised by the mass of the shell and its equation of state;
in Sec.\ \ref{sec:geodesics} we give a qualitative description of the geodesic
motion in DBH spacetimes which paves the way for the future interpretation of
our results; in Sec.\ \ref{sec:waves_in_DBH} we review the description of scalar
wave propagation on a DBH spacetime. Sec.\ \ref{sec:resonances} contains the
main results of the paper, that is, the calculation of the Regge pole spectrum of DBH's.
We reveal in particular that in contrast with the isolated black hole case, the
DBH resonance spectrum contains several branches which correspond to the extra
structure and complexity of the spacetime. The spectra are computed numerically
by adapting the continued fraction method originally developed by Leaver.
We supplement our analysis with a WKB calculation to verify our numerical results.
In Sec.\ \ref{sec:scattering}, we compute numerically the scattering cross section
of planar waves impinging on a DBH and show that characteristic oscillations
appear in some configurations. The cross sections are then described using the
RP spectrum identified in Sec.\ \ref{sec:resonances} via the CAM representation.
Finally we conclude with a discussion of our results and future prospects in
Sec.\ \ref{sec:conclusion}.
We use natural units $(c=G=\hbar=1)$ throughout the paper.

\section{Dirty Black Hole spacetime}\label{sec:DBH_spacetime}

Here we briefly review the Israel formalism \cite{Israel:1966rt} applied to our case where
there are two distinct Schwarzschild geometries:
\be
ds^2 = f_\pm(r) dt_\pm^2 - \frac{dr^2}{f_\pm(r)} - r^2 d\Omega^2
\ee
where $d\Omega^2 = d\theta^2+ \sin^2\theta d\phi^2$ is the line element on a unit
sphere, $f_\pm(r) = 1 - 2M_\pm/r$ is the Schwarzschild potential for each side of the
shell, and we have already applied the knowledge that the shell is at a given radius
$r=R_s(\tau)$, which implies that the angular and radial coordinates are the same on
each side. At this point, we are not assuming the wall is static, hence there is a different
local time coordinate on each side of the shell, which has, in general, a time dependent
trajectory given by $\left (t_\pm(\tau),R_s(\tau)\right)$ with $\tau$ the proper time on
the shell: $f_\pm {\dot t}_\pm^2 - \dot{R}_s^2/f_\pm = 1$.

The Israel junction conditions read
\be
\Delta K_{ab} - \Delta K h_{ab} = 8\pi S_{ab} =
Eu_au_b - P \left ( h_{ab} - u_au_b \right)
\ee
where $h_{ab} = g_{ab} + n_a n_b$ is the induced metric on the shell (with
$n_\pm^a = ( \dot{R}_s, \dot{t}_\pm )$ an outward pointing unit normal on each side
of the shell), $\Delta K_{ab} = K^+_{ab} - K^-_{ab}$ is the jump in extrinsic curvature across
the shell, and $S_{ab} = \int_-^+ T_{ab}$ is the energy momentum of the wall, here assumed
to take a perfect fluid form.

Inputting the form of the geometry into the Israel equations results in two independent
``cosmological'' equations
\be
\beal
\dot{R}_s^2 + 1 &=
(4\pi E)^2R_s^2 + \frac{(M_++M_-)}{R_s} + \frac{(M_+-M_-)^2}{(8\pi E)^2R_s^4}, \\
\dot{E} &+ 2 \frac{\dot{R}_s}{R_s} (E+P) =0.
\eeal
\ee

A static shell requires both $\dot{R}_s$ and $\ddot{R}_s$ to be zero, which places
constraints between the values of the mass and shell energy-momentum.
As is conventional, we assume an equation of state for the shell
\be
P=wE\;, \qquad \Rightarrow \qquad E = \frac{\rho_0}{4\pi R_s^{2(1+w)}}
\ee
which then gives
\be
\beal
M_\mp &= \frac{4w(1+2w)R_s - (1+4w)^2M_\pm}{f_\pm(R_s)(1+4w)^2},\\
\rho_0^2 &= \frac{4R_s^{4w}(2wR_s - (1+4w)M_\pm)^2}{f_\pm(R_s)(1+4w)^2}.
\eeal
\ee

For a given equation of state, the minimum value of $R_s$ is when the shell
has vanishing energy (and $M_+=M_-$ trivially)
\be
R_{s,min} = \frac{(1+4w)}{2w} M_\pm.
\label{minshell}
\ee
As $R_s$ increases, the shell gradually contributes more mass to the spacetime,
leading to a larger disparity between $M_+$ and $M_-$. (If $R_s<R_{s,min}$,
$M_+<M_-$, and the energy of the shell is negative.)

An important feature of a black hole spacetime of relevance to QNM's and RP's
is the \emph{light ring}, or the unstable null circular geodesic orbit at $r =3M$.
We now see three different possibilities for the shell location depending on
where it is situated with respect to the light-rings of the interior
and exterior Schwarzschild masses ($3M_\pm$).

(i) $R_s < 3M_-$: If $w>1/2$, i.e.\ the equation of state is stiffer than radiation,
then \eqref{minshell} shows that it is possible for the shell to lie `inside' the
light-ring of the black hole. Since the local Schwarzschild mass is $M_+$
outside the shell, this means that the $3M_-$ light-ring no longer exists, and
the spacetime will have only one light-ring at $r=3M_+$. Conversely, if
$w\leq1/2$, then \eqref{minshell} shows
that the shell can never lie inside the inner light-ring.

(ii) $3M_- \leq R_S \leq 3M_+$: This configuration, where the shell lies between the
light-rings, is possible for equations of state with $w\in(\frac{\sqrt{3}-1}{4},1]$, and
the spacetime will have both light-rings present.

(iii) $R_s>3M_+$: If the shell lies outside the light-ring of the exterior geometry, then
again there is only one light-ring in the spacetime, although it is now
\emph{inside} the shell at $3M_-$. Demanding positivity of $R_s-3M_+$ for
$R_s\geq R_{s,min}$ shows that this requires the equation of state parameter
$w<1/2$.

\begin{figure}
\centering
\includegraphics[width=0.45\textwidth]{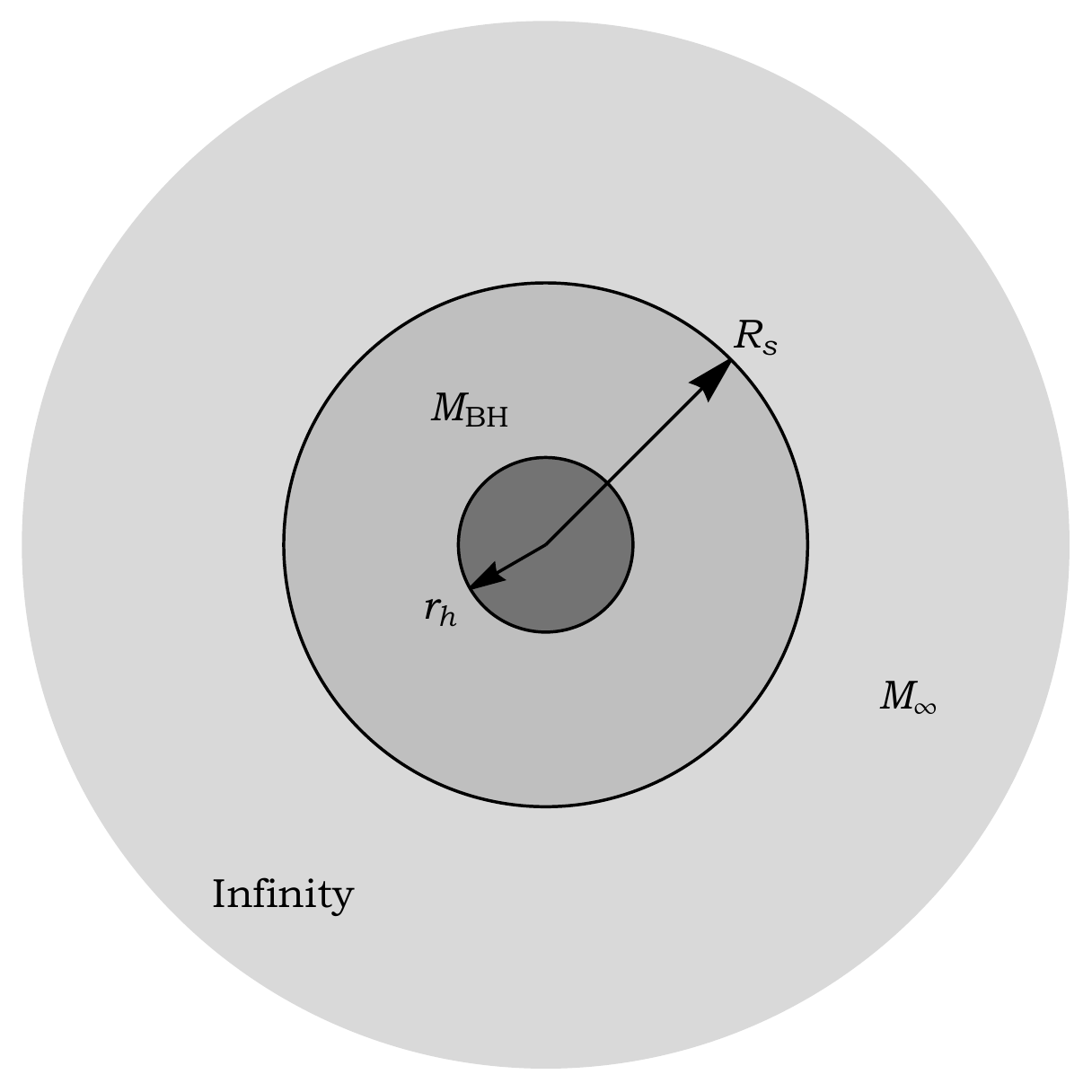}
\caption{Schematic diagram of a black hole surrounded by a thin shell
(the thickness of the shell is ignored) located at $r=R_s$.}
\label{fig:Schematic_diagram_DBH}
\end{figure}

Having derived all the possible configurations of the shell, we conclude this section
by altering our notation slightly to align with that of Macedo et al.\
\cite{Macedo:2015ikq}, in which a global coordinate system $(t,r,\theta,\phi)$
is used:
\begin{equation}\label{lineele}
ds^2 = A(r)dt^2-B(r)^{-1}dr^2-r^2d\Omega^2.
\end{equation}
Note, we did not begin with this global system as it is not possible to define a global
time coordinate unless the shell is static, and we wanted to derive all possibly consistent
static configurations commensurate with physical equations of state.
Replacing  $M_- = M_{\text{\tiny BH}}$ and $M_+ = M_\infty$,
these metric functions are:
\be
\beal
A(r) &= \begin{cases}
\alpha(1-2M_{\text{\tiny BH}}/r), & r < R_s \\
1-2M_{\infty}/r,                           & r > R_s,
\end{cases}
\\
B(r) &=
\begin{cases}
1-2M_{\text{\tiny BH}}/r, & r < R_s \\
1-2M_{\infty}/r,                     & r > R_s,
\end{cases}
\eeal
\label{AandB}
\end{equation}
where $\alpha$ is inserted to ensure the induced metric on the shell is well
defined from each side
\be
\alpha = \frac{R_s-2M_\infty}{R_s - 2M_{\text{\tiny BH}}} \; ,
\ee
and $B$ contains a jump at the shell to represent
the discontinuity in the extrinsic curvature.
A cartoon of the spacetime geometry is given in Fig.\ \ref{fig:Schematic_diagram_DBH}.

\section{Geodesic motion}\label{sec:geodesics}

DBH spacetimes can exhibit critical effects of geometrical optics.
We will see later that this rich geodesic structure
can be used to understand features in the scattering of waves presented in this paper.
The important notion emerging from the geodesic motion relevant to the study of
resonances is the \emph{light-ring} or \emph{photon sphere}.
It corresponds to a local maximum and an unstable equilibirum point of the
effective geodesic potential and
offers an intuitive link to QNMs~\cite{Cardoso:2008bp}.
It is therefore natural to ask about the structure of the light-rings in DBH spacetimes.
We note that the link between QNMs and light-rings is however not exact
\cite{Khanna:2016yow} and a simple geodesic analysis is not always sufficient to
characterise the resonances of a system, as we will see later.

Geodesic motion on static DBH spacetimes has been studied in, e.g.,
\cite{Macedo:2015ikq}, hence we simply note the key steps of the calculation.
The equation of motion of a null geodesic can, by a judicious choice of affine parameter,
be brought into the form
\be\label{eq:geo_pot}
\dot{r}^2 + B(r) \left ( \frac{L^2}{r^2} - \frac{1}{A(r)}\right)
= \dot{r}^2 + U_{\text{\tiny{eff}}}(r) = 0
\ee
where without loss of generality the geodesic is taken to lie in the equatorial plane,
$A$ and $B$ were given in \eqref{AandB}, and $L = r^2\dot{\phi}$ is a conserved
(not necessarily integer) quantity along the geodesic related to the angular momentum
(the affine parameter is chosen so that $A\dot{t}=1$).

The discontinuity of $B$ across the shell means this potential is also discontinuous,
however the interpretation of the level of the potential as a ``kinetic'' energy remains,
and we can use the graph of $U_{\text{\tiny{eff}}}(r) $ to not only interpret geodesic
motion, but also to infer more general scattering phenomena.
The main feature of the potential, resulting from the discontinuity, is the
possible existence of two local maxima.

\begin{figure}
\centering
\includegraphics[width=0.5\textwidth, trim = 0 0 0cm 0]{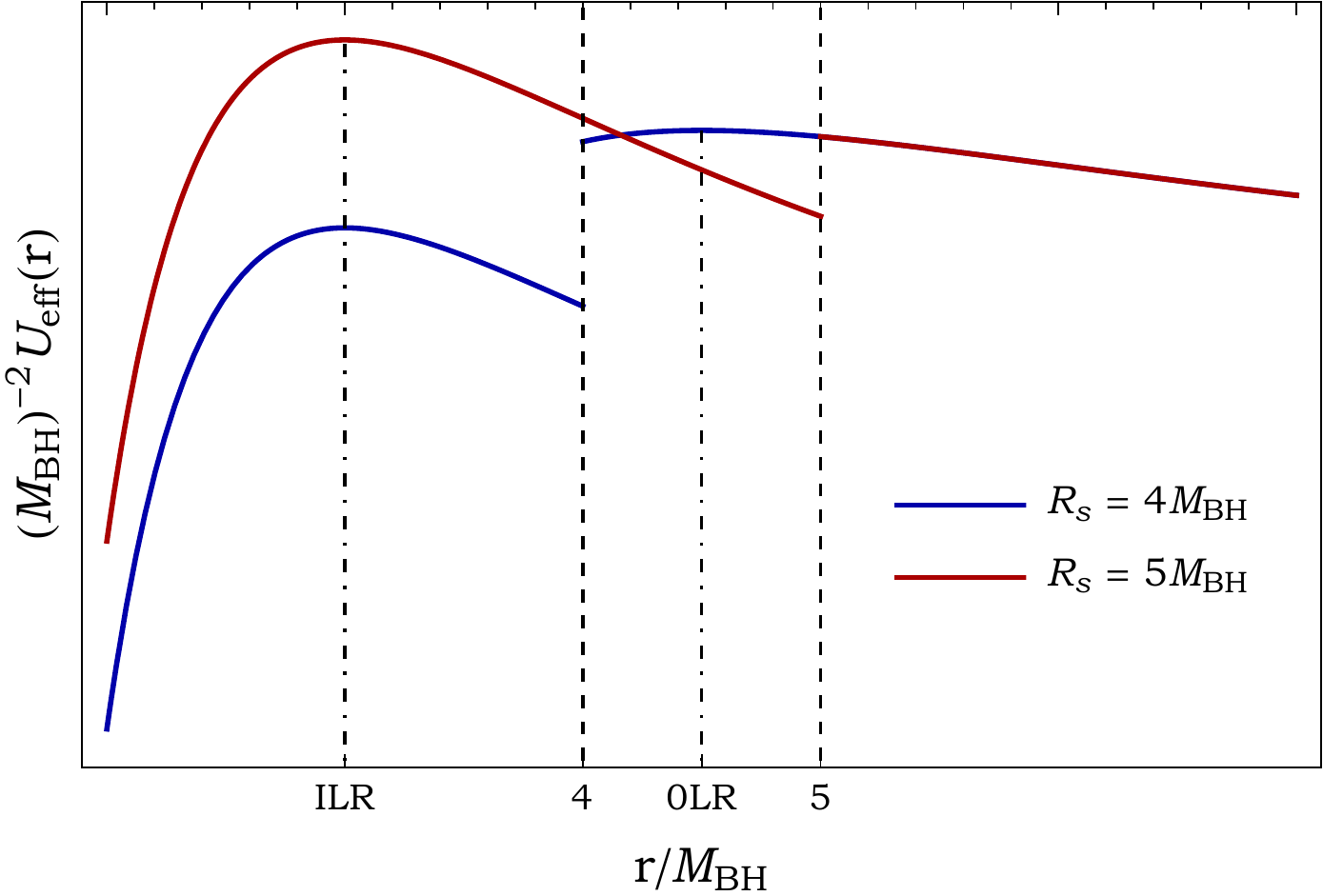}
\caption{Illustration of the geodesic potential for the
two configurations of DBH studied later on with $M=1.5M_{\rm{BH}}$ and
$R_S=4M_{\rm{BH}}$ (blue curve) or $R_S=5M_{\rm{BH}}$ (red curve).
In the latter case, the shell is located outside of the outer light-ring (OLR) and
therefore only exhibits a single light-ring structure. The former has the shell
between the OLR and inner light-ring (ILR) and exhibits the two light-rings structure.
(Features of the potential have been exaggerated to beter illustrate the structure.)}
\label{fig:geodesic_potential}
\end{figure}

A further interesting feature of these geometries is the possibility of geodesics
that remain trapped between the light-rings. While these obviously will not be
visible from far away, we might expect the existence of these resonant trajectories to
correspond to some imprint in the scattering cross sections. Indeed, even for
the case of a single light-ring, when that is located inside the shell, we see the
sharp discontinuity in the potential at the shell location gives rise to a sharp
``dip'' in the potential (see Fig.\ \ref{fig:geodesic_potential}).
While this dip results in a change of direction for the light ray, it does not
`trap' the (classical) geodesic, however a quantum mechanical system would
display an effect. The scattering of scalar waves of the dirty black hole might
therefore be expected to detect this dip.

As we will see in Sec.\ \ref{sec:resonances}, the properties of the local maxima
of the geodesic potential can be seen in the resonance spectrum leading to two
separated branches. We also identify a third branch related precisely to the dip
referred to above - indicative
of quasibound states trapped between the light-ring and the shell.

Before turning our attention to the spectrum of resonances of DBH, we first anticipate
the effects expected to be present in the scattering based on the geodesic structure
presented above. The link between geodesic motion and scattering effect is best
seen through the deflection angle which we now discuss.

\subsection{Deflection angle and classical scattering}
\begin{figure*}

\includegraphics[scale=0.5]{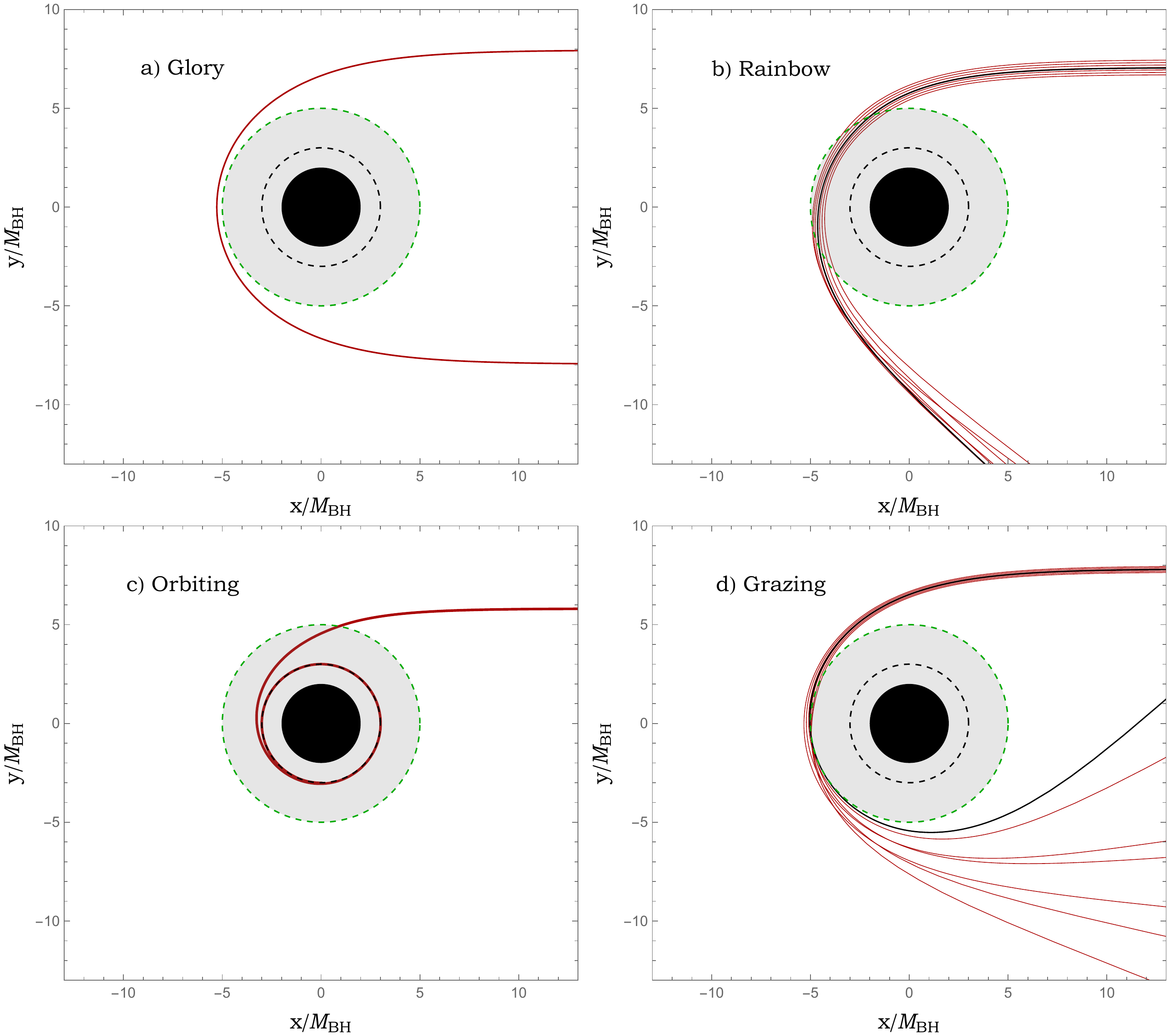}
\caption{Illustration of the various critical effects in a dirty black hole spacetime.
Here we have chosen the configuration where the shell lies outside the light-ring
of the exterior geometry, i.e., $R_S=5M_{\rm{BH}}$ and $M_{\infty}=1.5M_{\rm{BH}}$.
In all pictures, the black disc represents the inner black hole with mass
$M_{\rm{BH}}$, the dashed green circle represents the location of the shell at
$r=R_{S}$ and the dashed black circle depicts the inner light-ring at $r =3M_\text{BH}$.
a) The picture depicts glory scattering. In this case, the defection angle function
$\Theta(b)$  passes smoothly through $\pi$, i.e., geodesics are scattered
in the backward direction.
b) The picture illustrates rainbow scattering. In this case, a congruence of geodesics
centred around the rainbow impact parameter $b =b_r$ is presented (solid red curves).
The rainbow ray (black solid curve) defines an extremal angle, called the rainbow
angle, beyond which rays cannot be deflected locally.
c) The picture depicts the orbiting phenomenon. In this case, the deflection angle
is divergent for a critical impact parameter $b =b_c$ and a particle orbits indefinitely
at $r = 3M_\text{BH}$ around the light-ring.
d) The picture represents grazing rays. A congruence of rays centred around the
grazing impact parameter $b=b_{R_s}$ is shown (solid red curves). The grazing ray
(solid black curve) sets the boundary of the so-called edge region.}
\label{fig:critical_geo}
\end{figure*}

The deflection angle is an important geometrical quantity that allows us to analyze
in the classical limit the differential scattering cross section. In the case of DBH,
the classical scattering cross section for null geodesics can be defined as
\cite{FORD1959259} (see also \cite{Collins73})

\begin{equation}\label{classical_scatt_cross_sec}
\frac{d\sigma}{d\Omega} =\frac{b}{\sin(\theta)}\frac{1}{\Big{|}
\frac{d\Theta_\text{geo}(b)}{db}\Big{|}}
\end{equation}
and the geodesic deflection angle is given by
\begin{equation}
\label{Deflection_angle}
\Theta_\text{geo}(b) = 2\int_{0}^{u_0} du\left[B(u)\frac{1-A(u)b^2u^2}{A(u)b^2}\right]^{-1/2} - \pi,
\end{equation}
where $u=1/r$ and $u_0 = 1/r_0$ with $r_0$ being a turning point.

In Eqs.\ \eqref{classical_scatt_cross_sec}~and~\eqref{Deflection_angle},
the geodesic deflection angle is related to the scattering angle $\theta$ by
\begin{equation}\label{rela_def_angle_scatt_angle}
\Theta(b) +2n\pi =\pm \theta
\end{equation}
with $n \in \mathbb{N}$ such that $\theta$ remains in ``its interval of definition'',
i.e.\ $[0,\pi]$ and $b$ is the impact parameter of the scattered null geodesic.

There are four main scattering effects, each associated with a specific property of
the deflection angle that we briefly review for the reader before qualitatively describing
the scattering of waves in DBH spacetimes. They can be cast into two classes:
the divergent class or the interference class.

\subsubsection{The divergent class: glories and rainbows}

\textit{- The glory:} If the deflection function passes smoothly through $0$ or $\pi$,
i.e.\ if geodesics are scattered in the forward or backward direction, then the
semi-classical cross section contributions from these geodesics will diverge.
Expanding the deflection function around the glory point to linear order, one can
cure this singularity and obtain a semi-classical expression for the glory in terms
of Bessel functions \cite{ADAM2002229}.
The typical behaviour associated with the glory is therefore an increase of the
scattering amplitude in the forward or backward direction. The glory effect is well
known in black hole physics and also appears in these DBH spacetimes.
Fig.\ \ref{fig:critical_geo}\,a) shows one backscattered ray contributing to the
glory effect in the case of a DBH studied in the following.

\textit{- Rainbow scattering:} From \eqref{classical_scatt_cross_sec}, we can see that
$d\sigma/d\Omega $ can diverge for $\theta\neq 0$ or $\pi$ if $d\Theta_{geo}/db = 0$.
An impact parameter $b_r$, for which the deflection function is stationary defines
an extremal angle, called the rainbow angle $\theta_r$.
For $b \sim b_r$, no rays can be deflected beyond $\theta_r$.
This will result in interferences on one side of the rainbow angle (known as the
illuminated side) and to exponential decay on the other (known as the dark side).
Similarly to the glory, one can expand the deflection function to second order in the
vicinity of the rainbow point to obtain a semi-classical description built on the Airy function.
The typical behaviour of rainbow scattering is an enhanced amplitude near the
rainbow angle with oscillations on one side of the rainbow and exponential decay
on the other side. This effect was shown to appear in astrophysical settings
\cite{Dolan:2017rtj} as well as in gravitational analogues \cite{Torres:2022zua}.
Fig.\ \ref{fig:critical_geo}\,b) shows rainbow rays in the case of a DBH studied in the following.

\subsubsection{The interference class: orbiting and grazing}
\textit{- Orbiting:}
Orbiting belongs to the interference class of critical effects and is associated with
a critical impact parameter $b_c$ for which the deflection angle diverges
$\lim_{b\rightarrow b_c}\Theta_{geo} =\infty$. This implies that geodesics
with an impact parameter close to $b_c$ can be deflected with arbitrary angles.
In other words, they orbit around the scatterer. This divergence also implies that
there will be infinitely many geodesics deflected at any angles, causing
interference at arbitrary angles. Orbiting is well known in black hole physics
and is associated with the existence of the light-ring previously mentioned
\cite{Andersson:2000tf}. Fig.\ \ref{fig:critical_geo}\,c) shows the orbiting of
geodesics in a DBH spacetime.
Since orbiting allows geodesics to be deflected to arbitrary angles, it is naturally
linked to the glory effect. Indeed it was shown, using the CAM approach, that the
two effects can be incorporated in a unified semi-classical formula built around
the properties of surface waves propagating on the light-ring~\cite{Folacci:2019vtt}.

\textit{- Grazing:}
Grazing is another critical effect belonging to the interference class. Grazing
(also known as edge effects) appear when the scatterer exhibits a discontinuity
\cite{Nus92,Nussenzveig1664764}. The discontinuity of the scatterer will lead to
a singular point in the deflection function which defines an extremal angle beyond
which rays cannot be deflected. This is similar to the rainbow effect with the main
difference being the fact that the deflection function is not stationary at the
grazing angle. Therefore, grazing will lead to interferences but does not necessarily
imply an increase of the scattering amplitude as was the case in rainbow
scattering. Fig.\ \ref{fig:critical_geo}\,d) shows grazing rays in a DBH spacetime
studied in the following.

\subsection{Critical effects in DBH spacetimes}

Equipped with the terminology and intuition from the semi-classical
description of critical effects, we now turn our attention to scattering in DBH spacetimes.

\begin{figure}
\centering
\includegraphics[width=0.5\textwidth]{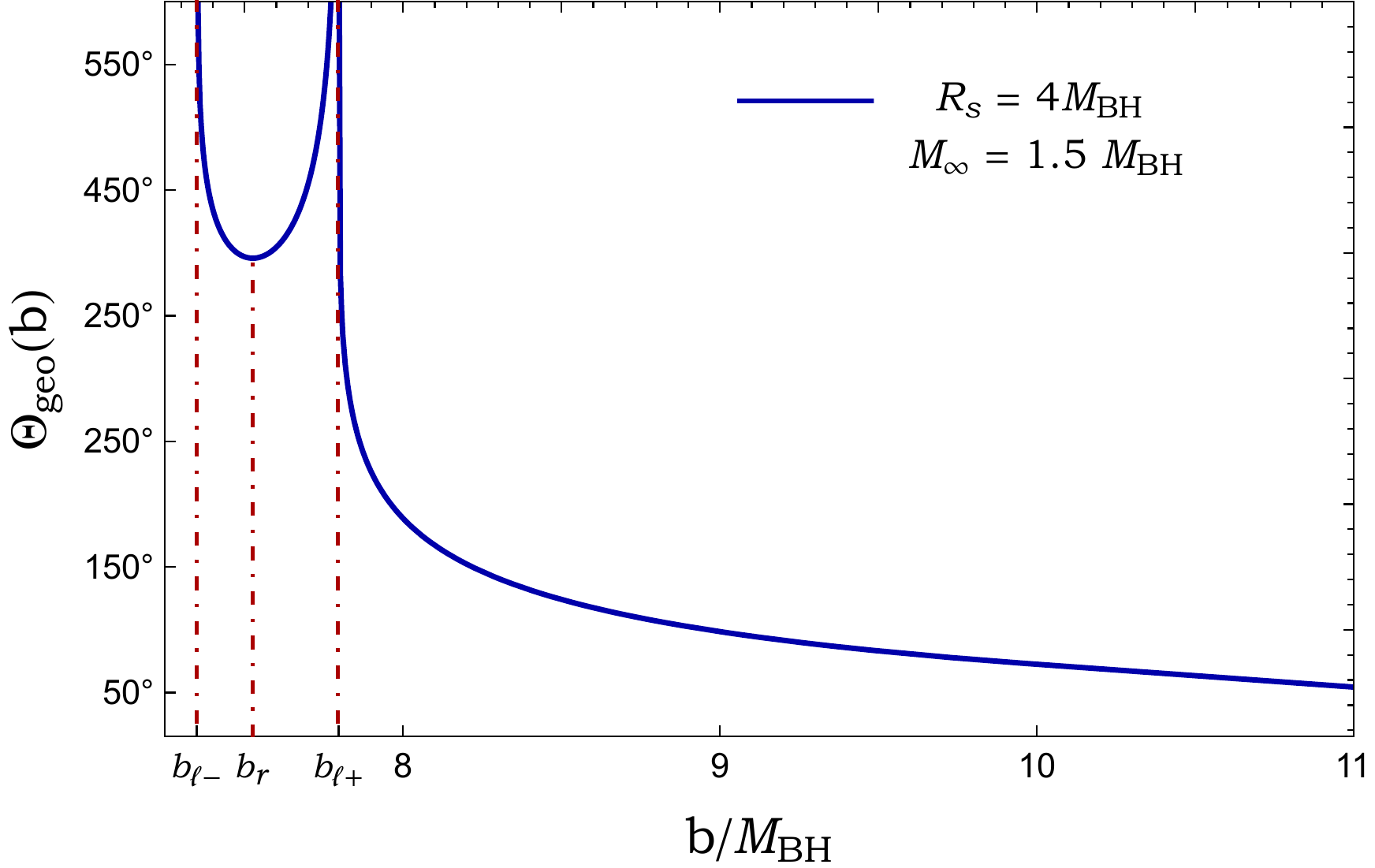}
\caption{Deflection angle as a function of the impact parameter obtained from
\eqref{Deflection_angle}. Here we assume $M_\infty = 1.5 M_\text{BH}$, and the
shell position $R_s = 4 M_\text{BH}$. The deflection angle diverges at both
critical parameters $b_{\ell_-} = 3 \sqrt{3} M_\text{BH}/\sqrt{\alpha}$ and
$b_{\ell_+} = 3\sqrt{3} M_\infty$ associated with a light-ring at $r_{\ell_-} = 3M_\text{BH}$
and $r_{\ell_+} = 3M_\infty$ respectively. There is also a stationary point in the
deflection angle, i.e.\ $\Theta^{'}_\text{geo}(b_r) = 0$, leading to the rainbow effect.}
\label{fig:deflection_angle_rs_4}
\end{figure}

\begin{figure}
\centering
\includegraphics[width=0.5\textwidth]{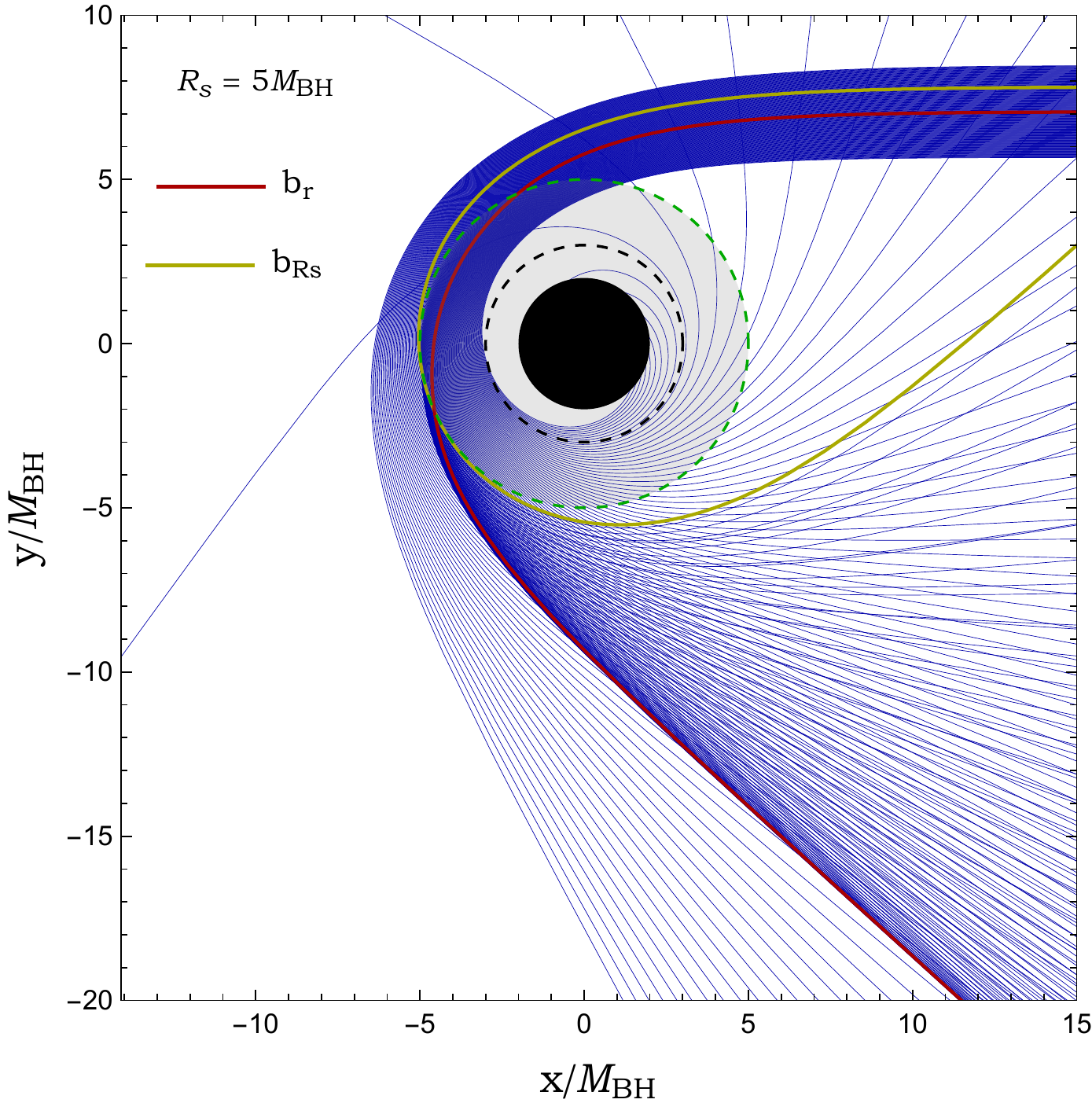}
\caption{Null geodesics scattered by a DBH. In this case, we assume
$M_\infty = 1.5 M_\text{BH}$ and $R_s = 5M_\text{BH}$ and the incident
geodesics are equally spaced. The impact parameter varies between
$0.95 b_c < b < 1.35 b_c$ with a fixed step size $\Delta b = 0.04 b_c$.
The red solid line is the rainbow ray and has an impact parameter
($b = b_r  = 1.1677 b_c$) and, we can see a high concentration of rays
in this direction leading to the characteristic amplification of the rainbow effect.
The olive curve represent the grazing ray.
}
\label{fig:ray_optic_Rs_5}
\end{figure}

As we noted in the previous section, there are three main geometric cases
classified by the location of the shell relative to the light-ring radii of the inner
and outer masses.
\\
\textbf{Case 1: $R_S<3M_{\rm{BH}}$.} In this case, the shell is located inside the
inner light-ring and therefore cannot be probed by geodesics escaping to infinity.
From the point of view of geodesics, this case is therefore similar to the one of
an isolated black hole.
\\
\textbf{Case 2: $3M_{\rm{BH}}<R_S<3M_{\infty}$.}
If the shell lies between the light-rings, then there is the possibility for geodesics to
probe each light-ring separately. Each light-ring is associated with a divergence of the
deflection angle, leading to orbiting. In the case
where both light-rings are accessible to geodesics, we expect two such divergences
to be present, associated with orbiting around the inner and outer light-rings.
Hence we further anticipate that there exists a value of $b_{\ell_-}<b<b_{\ell_+}$ such that
the deflection angle is extremal. This implies the presence of a fold caustic and
leads to rainbow scattering~\cite{FORD1959259}. The deflection angle revealing
these critical effects is represented in Fig.\ \ref{fig:deflection_angle_rs_4} for the
configuration where the shell is located between the two light-rings. In this
configuration, and for the choice of parameters $R_S=4M_{\rm{BH}}$ and
$M_{\infty}=1.5M_{\rm{BH}}$, the rainbow angle $\theta_r\approx 35.8^{\circ}$
and is associated with the impact parameter $b_r$.
The presence of rainbow
scattering in DBH spacetimes was noted in~\cite{Leite:2019uql} but its characteristic amplification was not clearly observed in their simulation due to the specific choice of DBH configurations.
\\
\begin{figure}
\centering
\includegraphics[width=0.5\textwidth]{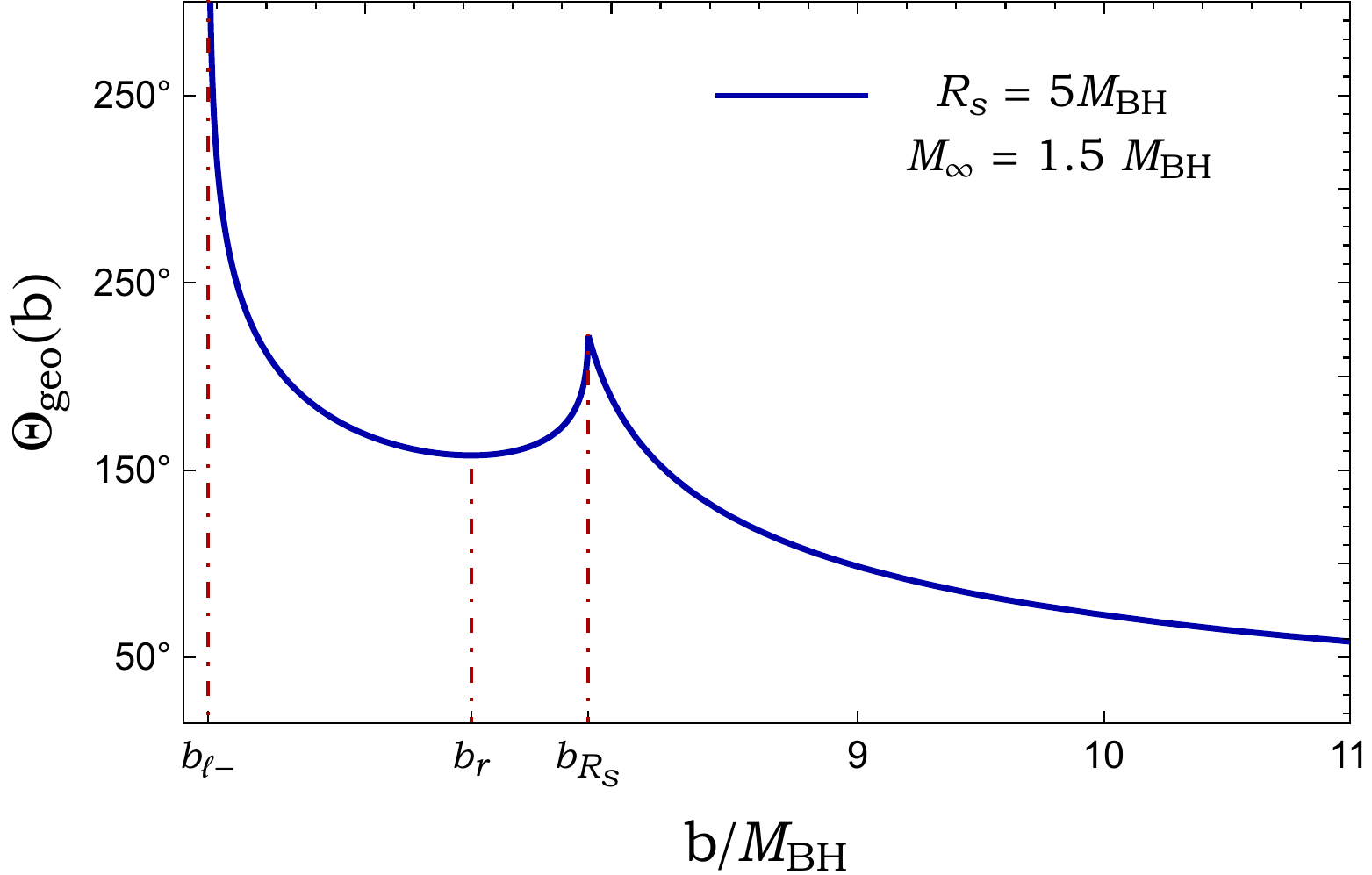}
\caption{Deflection angle as a function of the impact parameter obtained from \eqref{Deflection_angle}. Here we assume $M_\infty = 1.5 M_\text{BH}$, and the
shell position $R_s = 5 M_\text{BH}$. The deflection angle diverges at a critical
parameter $b_{\ell_-} = 3 \sqrt{3} M_\text{BH}/\sqrt{\alpha}$ associated with a
light-ring at $r_{\ell_-} =3M_\text{BH}$. There is also a stationary point in the
deflection angle, i.e.\ $\Theta^{'}_\text{geo}(b_r) = 0$, leading to the rainbow effect.
A local maximum occurring at $b = b_{R_s}=\sqrt{R_S^3/(R_S-2M_\infty)}$ is also
present, leading to grazing.
A congruence of geodesics in this specific DBH configuration is represented in
Fig.\ \ref{fig:ray_optic_Rs_5}.}
\label{fig:deflection_angle_rs_5}
\end{figure}
\\
\textbf{Case 3: $R_S  > 3M_\infty$}.
If the shell lies outside $3M_\infty$, a single light-ring is present in the geometry,
which is now located \emph{inside} the shell.
In this case, the orbiting effect is qualitatively the same as for an isolated black hole
and is associated with the critical impact parameter $b = 3\sqrt{3}M_\text{BH}/\sqrt{\alpha}$.
In this configuration, a second local maximum of the potential is present. Contrary to
the previous case, this maximum is not associated with the second light-ring but
rather to the presence of the shell (see Fig.\ \ref{fig:geodesic_potential}).
Since this second local maximum is not an (unstable) equilibrium point, it will not be
associated with a divergence of the deflection angle but rather with a singular point
(i.e.\ the deflection function is not differentiable).
It is important to note that the singular point occurring at $b=b_{R_S}
= \sqrt{R_S^3/(R_S-2M_\infty)}$ is a local
maximum of the deflection angle, but does not satisfy the condition
$d\Theta_{geo}/db = 0$, which implies that it is \emph{not} associated to rainbow
scattering as stated in~\cite{Leite:2019uql}. Instead, this singular point will lead
to grazing as discussed previously. Similar to the previous case, there is an impact
parameter associated to rainbow scattering between the orbiting impact parameter
and the grazing one. In this configuration, and for the choice of
parameters $R_S=5M_{\rm{BH}}$ and $M_{\infty}=1.5M_{\rm{BH}}$, the
rainbow angle $\theta_r\approx 157.8^{\circ}$.
Fig.\ \ref{fig:deflection_angle_rs_5} depicts the deflection function of such a
configuration and Fig.\ \ref{fig:ray_optic_Rs_5} depicts a congruence of
geodesics in this DBH set-up. The red and dark green rays in Fig.\
\ref{fig:ray_optic_Rs_5} represent the rainbow and grazing ray respectively
and are both associated with an extremal angle. Note the qualitative difference
between the rainbow and grazing ray which resides in the concentration of
geodesics around each critical trajectory.

\section{Waves on a dirty black hole spacetime}\label{sec:waves_in_DBH}

We consider a scalar field, $ \Phi(x)$, propagating on the DBH spacetime,
governed by the Klein-Gordon equation
\begin{equation}
\Box \Phi \equiv \frac{1}{\sqrt{-g}} \partial_{\mu} \left( \sqrt{-g}
g^{\mu \nu} \partial_{\nu} \Phi \right) = 0
\end{equation}
where $g^{\mu \nu}$ is the inverse metric and $g$ is the metric determinant.
Performing a standard separation of variables,
\begin{equation}
\Phi = \frac{1}{r} \sum_{\omega \ell m} \phi_{\omega \ell}(r)
Y_{\ell m}(\theta, \phi) e^{-i \omega t} ,
\label{eq:sepvariables}
\end{equation}
leads to a radial equation of the form
\begin{equation}
\label{H_Radial_equation}
\left[\frac{d^{2}}{dr_{\ast}^{2}}+\omega^{2}-V_{\ell}(r)\right]\phi_{\omega\ell}= 0,
\end{equation}
where $V_{\ell}(r)$ is the effective potential which is given by
\begin{equation}
\label{Potential}
\beal
V_{\ell}(r) = &A(r) \Bigl[ \frac{\ell(\ell+1)}{r^2}\\
&+\frac{2}{r^3} \left ( M_{\text{\tiny BH}} \Theta(R_s-r) + M_\infty \Theta(r-R_s)
\right) \Bigr]
\eeal
\ee
with $\Theta$ being the Heaviside step function,
and $r_{\ast}$ denotes the \emph{tortoise coordinate} defined by
\begin{equation}
\label{tortoise_coord}
  \frac{dr_*}{dr} = \frac{1}{\sqrt{A(r)B(r)}},
\end{equation}
\textit{viz.}
\begin{equation}
\label{tortoise_coordinate}
r_{\ast}(r) =
\left\{
\begin{aligned}
&\frac{1}{\sqrt{\alpha}}\left[r +M_{\text{\tiny BH}}\ln\left(\frac{r}{M_{\text{\tiny BH}}}
-1\right)\right]+ \kappa & \scriptstyle{M_{\text{\tiny BH}} < r \leq R_s,}\\
& r + 2M_\infty\ln\left(\frac{r}{2M_\infty}-1\right)  & \scriptstyle{r > R_s,}
\end{aligned}
\right.
\end{equation}
where the constant $\kappa$ is fixed so that $r_{\ast}(r)$ is continuous.

In the following, we introduce the modes $\phi_{\omega\ell}^{\text{in}}$ which are
solutions of \eqref{H_Radial_equation} and are defined by their behaviour at the
horizon $r =2M_{\text{\tiny BH}}$ (i.e., for $r_* \to -\infty$) and at spatial infinity
$r \to +\infty$ (i.e., for $r_* \to +\infty$):
\begin{eqnarray}
\label{bc_in}
& & \phi_{\omega \ell}^{\mathrm {in}}(r_{*}) \sim \left\{
\begin{aligned}
&\displaystyle{e^{-i\omega r_\ast}}  & \, (r_\ast \to -\infty),\\
&\displaystyle{A^{(-)}_\ell (\omega) e^{-i\omega r_\ast} + A^{(+)}_\ell (\omega) e^{+i\omega r_\ast}} & (r_\ast \to +\infty).
\end{aligned}
\right. \nonumber\\
& &
\end{eqnarray}
Here, the coefficients $A^{(-)}_\ell (\omega)$ and  $A^{(+)}_\ell (\omega)$
appearing in \eqref{bc_in} are complex amplitudes and allow us to define the
scattering $S$-matrix elements,
\begin{equation}\label{Matrix_S}
  S_{\ell}(\omega) =  e^{i(\ell+1)\pi} \, \frac{A_{\ell}^{(+)}(\omega)}{A_{\ell}^{(-)}(\omega)}.
\end{equation}

It should be noted that, due to the choice of coordinates (keeping $r$ as the areal
radius on both sides of the shell) the transverse metric is discontinuous at the shell
meaning that we have to also place a boundary condition on the eigenfunctions
$\phi_{\omega\ell}$ at the shell \cite{Macedo:2015ikq}:
\be
\beal
\bigl[ &\sqrt{B(R_s)} \left (R_s \phi_{\omega \ell}^\prime(R_s)
- \phi_{\omega \ell}(R_s) \right )\bigr]_{+} \\
&\qquad = \left[ \sqrt{B(R_s)} \left (R_s \phi_{\omega \ell}^\prime(R_s)
- \phi_{\omega \ell}(R_s) \right )\right]_{-}
\eeal
\label{jump_condition}
\ee

\section{Resonances of the dirty black holes}
\label{sec:resonances}

\subsection{Quasinormal modes and Regge poles}

Resonant modes are characteristic solutions of the wave equation,
\eqref{H_Radial_equation}, satisfying purely ingoing/outgoing boundary
conditions at the horizon/infinity.
Their spectrum is the set of zeros of the scattering matrix $S_{\ell}(\omega)$
\eqref{Matrix_S}, i.e., a simple pole of $A_\ell^{(-)}(\omega)$.
It can be seen as a set of frequencies $\omega_{\ell n}$ in the complex-$\omega$
plane at which the scattering matrix $S_{\ell}(\omega)$ has a simple pole for
$\ell \in \mathbb{N}$ and $\omega_{\ell n} \in \mathbb{C}$ (the so-called
\textit{quasinormal mode spectrum}), or a set of angular momenta
$\lambda_n(\omega)\equiv \ell_n+1/2$ in the complex-$\lambda$ plane
at which the scattering matrix has a simple pole for $\omega \in \mathbb{R}$
and $\lambda_{n}(\omega) \in \mathbb{C}$ (the so-called \textit{Regge pole spectrum}).
Here $n = 1,2,3,\ldots $ labels the different elements of the spectrum and is
referred to as the overtone number.

The quasinormal mode spectrum of DBHs has been investigated for configurations
where $M_\infty - M_{\text{\tiny BH}}\ll 1$ using perturbative techniques
\cite{Leung:1997was,Leung:1999iq}. We shall now consider, for the first time,
the Regge poles of generic DBHs.

\subsection{Numerical method}

To compute the QNM/Regge pole spectrum of dirty black holes, we follow the method
of Ould El Hadj et al.\ \cite{OuldElHadj:2019kji} who calculated the Regge pole spectrum
of scalar and gravitational waves (in the axial sector) for a gravitating compact body.
Their method is an extension of the original continued fraction method developed
originally by Leaver~\cite{Leaver:1985ax,leaver1986solutions}.

The method involves writing the solution to the wave equation
\eqref{H_Radial_equation} as a power series around a point $b$ located
outside the shell,
\begin{equation}\label{ansatz_CFM}
\phi_{\omega,l}(r) = e^{i\omega r_*(r)} \sum_{n=0}^{+\infty} a_n \left(1 - \frac{b}{r}\right)^n\,,
\end{equation}
where the coefficients $a_n$ obey a four-term recurrence relation:
\begin{equation}\label{Recurrence_4_terms}
\alpha_n a_{n+1} + \beta_n a_{n} +\gamma_{n} a_{n-1} +\delta_{n} a_{n-2}  = 0,
\quad \forall n\geq 2 ,
\end{equation}
where
\begin{subequations}
\begin{eqnarray}\label{Coeffs_3_termes}
&& \alpha_n = n (n+1)\left(\!1-\frac{2M_\infty}{b}\!\right),    \\
&& \beta_n  = n\left[\left(\!\frac{6M_\infty}{b}-2\!\right)n + 2ib\omega\right] ,  \\
&& \gamma_n = \left(\!1-\frac{6M_\infty}{b}\!\right)n(n-1)-\frac{2M_\infty}{b}-\ell(\ell+1) ,  \\
&& \delta_n = \left(\!\frac{2M_\infty}{b}\!\right)\left(n-1\right)^2 .
\end{eqnarray}
\end{subequations}
The initialisation coefficients, $a_0$ and $a_1$, are found directly from
\eqref{ansatz_CFM},
\begin{eqnarray}\label{Initial_Conds}
&& a_0 = e^{-i\omega r_*(b)}\phi_{\omega_\ell}(b) ,  \\
&& a_1 = b e^{-i\omega r_*(b)}\left(\frac{d}{dr}\phi_{\omega_\ell}(r)
\Big{|}_{r=b}-\frac{i\omega b}{b-2M_\infty}\phi_{\omega_\ell}(b)\right).
\end{eqnarray}
In practise, the coefficients $a_0$ and $a_1$ are found numerically by integrating
\eqref{H_Radial_equation} from the horizon up to $r = b> R_s$.

In order to apply Leaver's method, we first perform a Gaussian elimination step in
order to reduce the 4-term recurrence relation to a 3-term recurrence relation:
\begin{equation}
\hat{\alpha}_n a_{n+1} + \hat{\beta}_n a_n + \hat{\gamma}_n a_{n-1} = 0,
\end{equation}
where we have defined the new coefficients, for $n\geq 2$:
\begin{subequations}
\begin{eqnarray}
&&\hat{\alpha}_n  = \alpha_n, \\
&& \hat{\beta}_n  = \beta_n - \hat{\alpha}_{n-1}\frac{\delta_n}{\hat{\gamma}_{n-1}},
\ \text{and} \\
&& \hat{\gamma}_n = \gamma_n - \hat{\beta}_{n-1}\frac{\delta_n}{\hat{\gamma}_{n-1}}.
\end{eqnarray}
\end{subequations}
The series expansion \eqref{ansatz_CFM} is convergent outside the shell provided
that $a_n$ is a minimal solution to the recurrence relation and $b/2<R_s<b$
\cite{Benhar:1998au}.
The existence of a minimal solution implies that the following continued fraction holds:
\begin{equation}\label{eq:CF}
\frac{a_1}{a_0} = \frac{-\hat{\gamma}_1}{\hat{\beta}_1 -}\,
\frac{\hat{\gamma}_2 \hat{\alpha}_1}{\hat{\beta}_2 -} \,
\frac{\hat{\gamma}_3 \hat{\alpha}_2}{\hat{\beta}_3 -} ...
\end{equation}
The above relation (or any of its inversions) is the equation, written in the standard form of continued fractions, we are solving in
order to find the RP/QNM spectrum.
In practise, we fix $\omega$ (equivalently $\ell$), and define a function
$f(\ell\in \mathbb{C})$ (equivalently $f(\omega \in \mathbb{C})$), which
gives the difference between the left-hand side and right-hand side of the
condition \eqref{eq:CF}. We then find the zeros of the function $f$ starting
from an initial guess.

An alternative method to the continued fraction expression is to use the Hill
determinant~\cite{mp}, where one looks for the zeros of the following determinant
\begin{equation}
\label{Determinant_Hill_4_termes}
D   =  \begin{vmatrix}
\beta_0 &  \alpha_0 &  0 & 0 & 0 &  \ldots  & \ldots  &  \ldots \\
\gamma_1 & \beta_1 & \alpha_1 & 0  &  0  &  \ldots  &  \ldots  &  \ldots  \\
\delta_2 &  \gamma_2 &  \beta_2  &  \alpha_2 &  0  &  \ldots  &  \ldots  &  \ldots  \\
\vdots & \ddots & \ddots  &  \ddots  & \ddots  &  \ddots  &  \ldots  &  \ldots \\
\vdots & \vdots & \delta_{n-1} & \gamma_{n-1} & \beta_{n-1} & \alpha_{n-1} & \ddots & \ldots \\
\vdots & \vdots & \vdots & \delta_n & \gamma_{n} & \beta_{n} & \alpha_{n} & \ddots   \\
\vdots & \vdots & \vdots & \vdots & \ddots & \ddots & \ddots & \ddots
\end{vmatrix} = 0.
\end{equation}
By assuming $D_n$
\begin{equation}
\label{derecurrence_4_termes}
D_n=\beta_n D_{n-1} - \gamma_{n}\alpha_{n-1}D_{n-2}
+ \delta_n \alpha_{n-1} \alpha_{n-2} D_{n-3} ,
\end{equation}
to be the determinant of the $n \times n$ submatrix of $D$
with the initial conditions
\begin{equation}
\label{Determinant_initial_conds}
\begin{split}
D_0 &=\beta_0, \\
D_1 &=\beta_1\beta_0-\gamma_1\alpha_0 , \\
D_2 &=\beta_0(\beta_1 \beta_2 - \alpha_1 \gamma_2)
- \alpha_0(\alpha_1 \delta_2 -\gamma_1 \beta_2) .
\end{split}
\end{equation}
the Regge poles (QNM frequencies) are found by solving  numerically
the roots $\lambda_n$ ($\omega_n$) of $D_n$.

We note that in order to confirm our results, we have used both methods and
they give the same results, up to the numerical precision.

\subsection{Results: The Regge pole spectrum}

In order to check the robustness of our numerical code, we first compute the
QNM spectrum for a configuration of DBH considered by Leung
et al.\ \cite{Leung:1997was,Leung:1999iq}, i.e., for
$M_\infty  = 1.02 M_{\text{\tiny BH}}$, $R_s = 2,52 M_{\text{\tiny BH}}$
and assuming $M_{\text{\tiny BH}} =1/2$.

\begin{figure}[htb]
\centering
\includegraphics[scale=0.50]{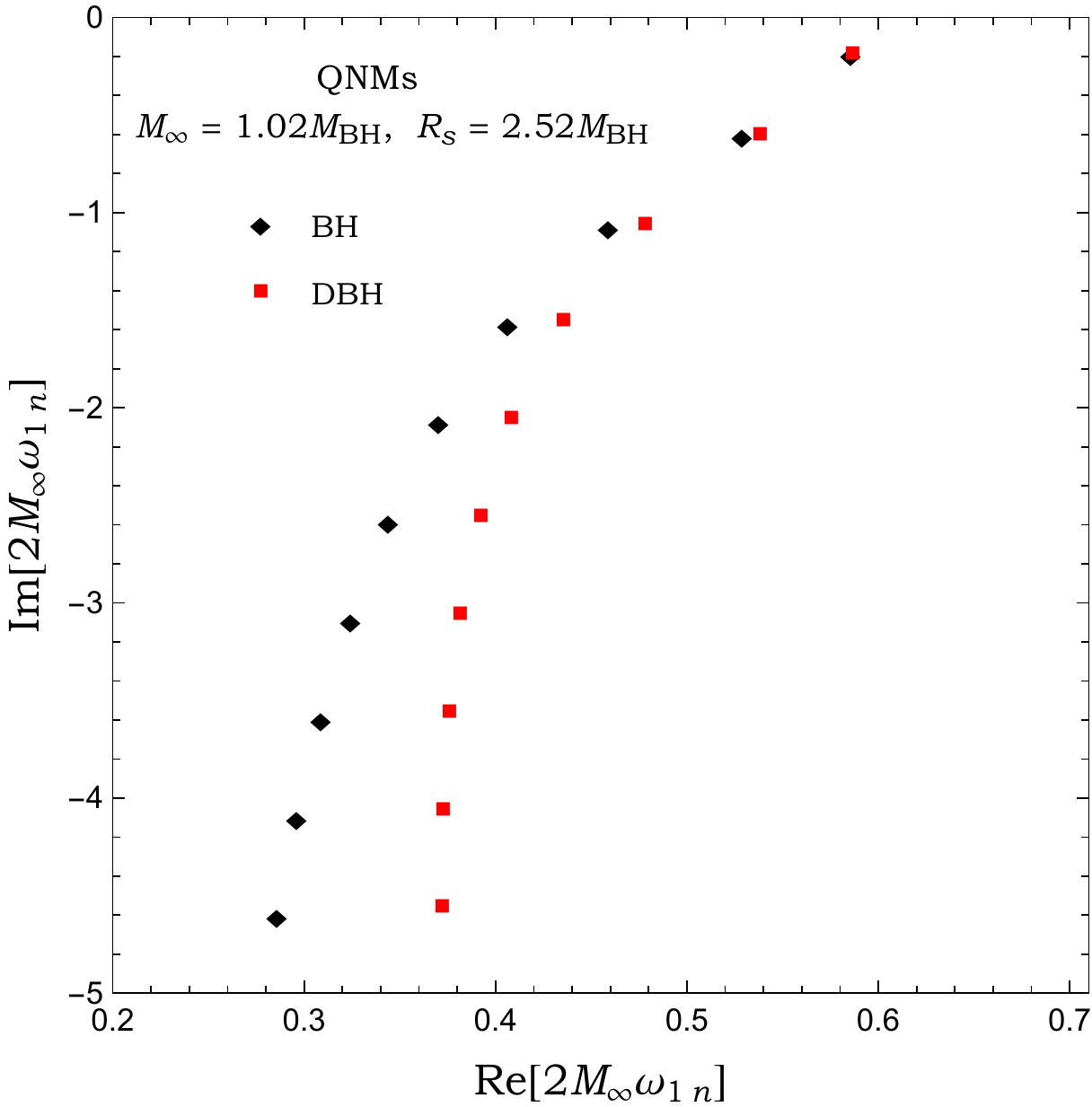}
\caption{\label{fig:QNMs_Plot_DBH_n_1_10} The ($\ell = 1, n = 1,\ldots,10$)
quasinormal modes of the scalar field. The results agree with Fig.\ 4 of
\cite{Leung:1999iq}. We assume $2M_{\text{\tiny BH}} = 1$.}
\label{fig:rp1}
\end{figure}

\begingroup
\begin{table}[htp]
\caption{\label{tab:table1} A sample of the first quasinormal frequencies
$\omega_{\ell n}$ of the scalar field. The radius of the thin shell is
\mbox{$R_s = 2.52M_{\text{\tiny BH}}$} and the ADM mass is $M_\infty = 1.02 M_{BH}$.
We assume $2M_{\text{\tiny BH}}=1/2$.}
\smallskip
\centering
\begin{ruledtabular}
\begin{tabular}{ccc}
$\ell$  & $n$ & $2M_\infty\omega_{\ell n}$ \\
\hline
$1$  & $1$  & $0.586628 - 0.190512 i$  \\[-1ex]
& $2$  & $0.538645 - 0.601030 i$  \\[-1ex]
& $3$  & $0.478474 - 1.064024 i$  \\[-1ex]
& $4$  & $0.435656 - 1.555056 i$   \\[-1ex]
& $5$  & $0.408698 - 2.055042 i$   \\[-1ex]
& $6$  & $0.392365 - 2.556656 i $  \\[-1ex]
& $7$  & $0.381985 - 3.058495 i$   \\[-1ex]
& $8$  & $0.376255 - 3.559562 i $  \\[-1ex]
& $9$  & $0.372953 - 4.060316 i $  \\[-1ex]
& $10$ & $0.372266 - 4.560224 i$   \\
\end{tabular}
\end{ruledtabular}
\end{table}
\endgroup
In Fig.\ \ref{fig:QNMs_Plot_DBH_n_1_10}, we show the QNM spectrum
corresponding to the DBH configuration studied by Leung et al.\ for
($\ell =1, n = 1,\ldots, 10$). We see that it agrees with Fig.\ 4 of~\cite{Leung:1999iq}.
The data for quasinormal frequencies $\omega_{\ell n}$ is listed in
Table~\ref{tab:table1}.

In Figs~\ref{fig:PRs_2Mw_3_2Mw_6_Rs_4}, \ref{fig:PRs_2Mw_16_2Mw_32_Rs_4}
and \ref{fig:PRs_2Mw_16_2Mw_32_Rs_5}, we present the numerical results
for the Regge pole spectrum of DBH's in two configurations: (i) a DBH where the
shell is located between the inner and outer light-rings ($3M_\text{\tiny BH} < R_s
< 3M_\infty$), so that both light-rings are present in the geometry, and (ii) a DBH
where the shell is located outside the outer light-ring ($R_s > 3M_\infty$), so that
only a single light-ring is present in the geometry, but this is \emph{inside} the
shell. The Regge poles for each configuration are presented for various frequencies.

Fig.\ \ref{fig:PRs_2Mw_3_2Mw_6_Rs_4} shows the Regge pole spectrum for the
first configuration with parameters $M_\infty = 1.5M_{\text{\tiny BH}}$ and
\mbox{$R_s = 4M_{\text{\tiny BH}}$} for two different frequencies $2M_\infty\omega = 3$
and $6$. For both frequencies, the Regge pole spectrum exhibits two branches,
represented in blue and red. For low overtones, the two branches
merge, with the distinction becoming clear at higher overtones.
Note that the splitting occurs for smaller values of $n$ as one increases the
frequency.
We can identify the origin of both branches as coming from the existence of two
light-rings if $3M_{\rm{BH}}<R_S<3M_\infty$ or from the inner-ring and the shell
if $R_S > 3M_{\infty}$. This can be seen from the low overtone origin of the branches.
Indeed for the fundamental modes, their Regge poles can be estimated from
the critical impact parameters
$b_c=b_{\ell_\pm}$ or $b_c = b_{R_S}$, i.e., $\text{Re}(\lambda_n(\omega))
\sim \omega b_{c}$~\cite{PhysRevD.81.024031}
with $b_{\ell_+} = 3\sqrt{3} M_\infty$, $b_{\ell_-} = 3\sqrt{3}
M_{_\text{\tiny BH}}/\sqrt{\alpha}$ and $b_{R_S} = \sqrt{R_S^3/(R_S-2M_\infty)}$.
For example, for the configuration with $M_{\infty} = 1.5M_{\text{\tiny BH}}$,
$R_S = 5M_{\text{\tiny BH}}$ and $2M_{\infty}\omega = 32$, and we find that the real
part of the fundamental Regge pole associated with the inner photon sphere and
the shell are approximately $66.88$ and $84.33$ respectively which agree remarkably
well with the numerical values presented in Table~\ref{tab:table3}.

\begin{figure}[htb]
\centering
 \includegraphics[scale=0.50]{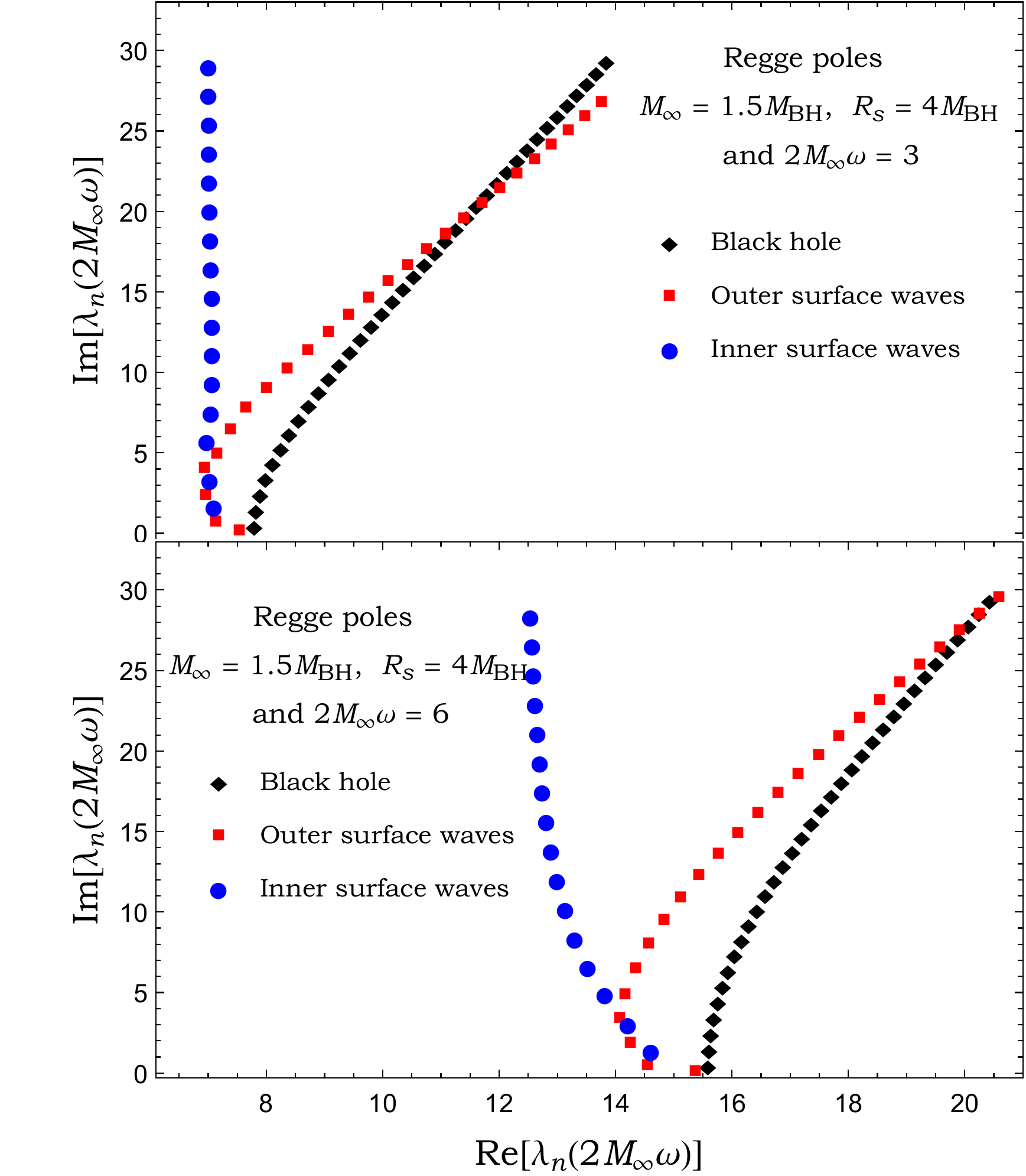}
\caption{\label{fig:PRs_2Mw_3_2Mw_6_Rs_4} The Regge poles
$\lambda_n(\omega)$ for the scalar field in a DBH spacetime with parameters
$M_\infty =1.5M_{\rm{BH}}$, $R_S = 4M_{\rm{BH}}$ and for frequencies
$2M_{\infty}\omega = 3$ (upper panel) and $2M_{\infty}\omega = 6$ (lower panel).
(We take $2M_{\text{\tiny BH}} = 1$ to produce these plots). In both panels, the
blue circle and red square branches correspond to the outer and inner surface
waves for the DBH spacetime while the black diamond branch is the one for an
isolated black hole of mass $M_{\text{\tiny BH}}$.}
\label{fig:rp1}
\end{figure}

\begin{figure}[htb]
\centering
 \includegraphics[scale=0.50]{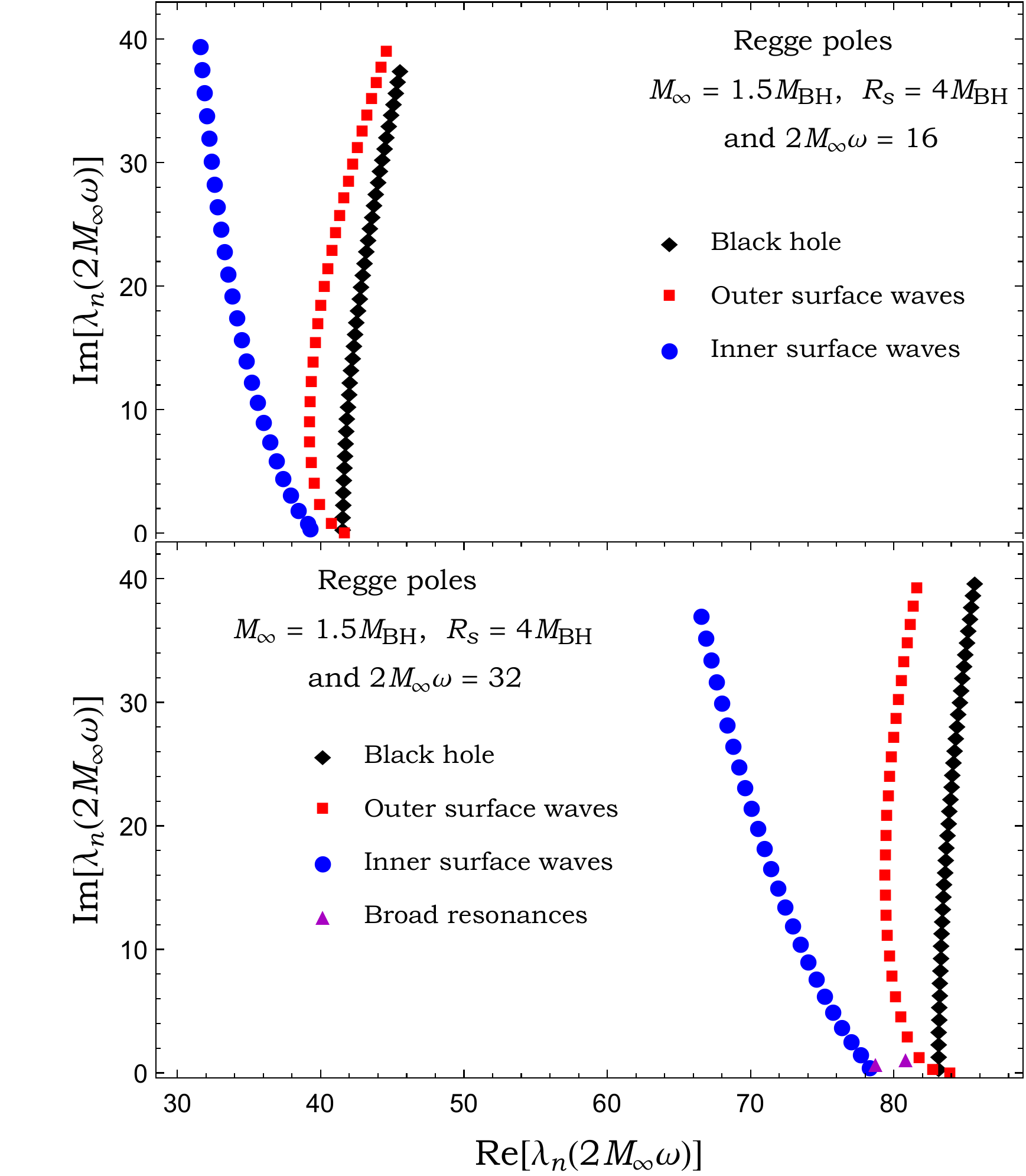}
\caption{\label{fig:PRs_2Mw_16_2Mw_32_Rs_4} The Regge poles $\lambda_n(\omega)$
for the scalar field in the DBH spacetime with parameters $M_\infty =1.5M_{\rm{BH}}$
and $R_s = 4M_{\rm{BH}}$ at frequencies $2M_{\infty}\omega = 16$ (upper panel)
and $2M_{\infty}\omega = 32$ (lower panel). We assume $2M_{\text{\tiny BH}} = 1$.
In both panels, the blue circle and red square branches correspond to the outer and
inner surface waves for the DBH spacetime while the black diamond branch is the
one for an isolated black hole. The purple triangles in the lower panel depict the
third branch of broad resonances.}
\label{fig:rp1}
\end{figure}

\begin{figure}[htb]
\centering
 \includegraphics[scale=0.50]{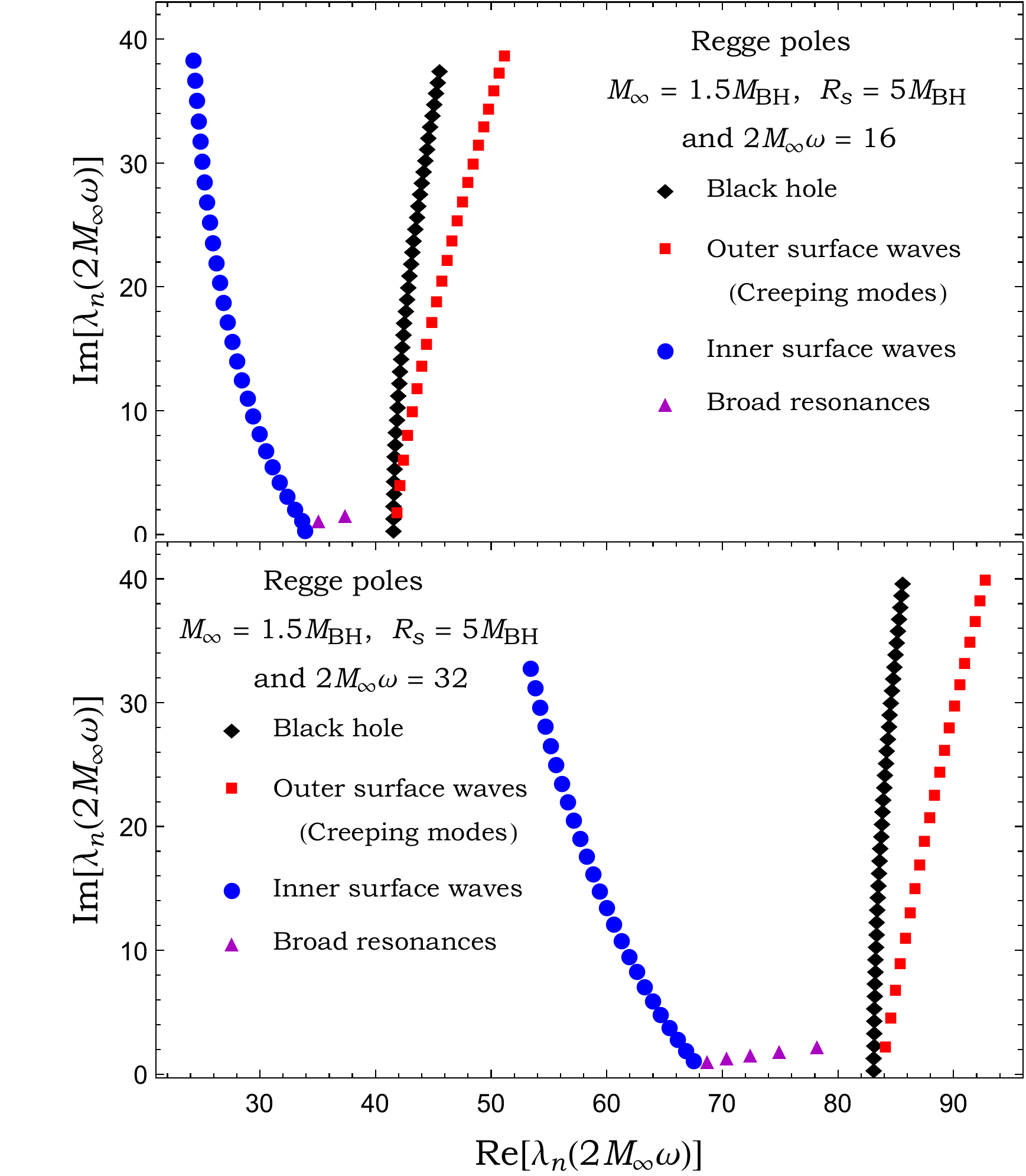}
\caption{\label{fig:PRs_2Mw_16_2Mw_32_Rs_5} The Regge poles $\lambda_n(\omega)$
for the scalar field in the DBH spacetime with parameters $M_\infty =1.5M_{\rm{BH}}$
and $R_s = 5M_{\rm{BH}}$ at frequencies $2M_{\infty}\omega = 16$ (upper panel)
and $2M_{\infty}\omega = 32$ (lower panel). We assume $2M_{\text{\tiny BH}} = 1$.
In both panels, the blue circle and red square branches correspond to the outer (creeping modes) and
inner surface waves for the DBH spacetime while the black diamond branch is the
one for an isolated black hole. The purple triangles depict the
third branch of broad resonances. As the frequency increases the surface wave
branches move away and more broad resonances appear in between them.}
\label{fig:rp2}
\end{figure}

Fig.\ \ref{fig:PRs_2Mw_16_2Mw_32_Rs_4} shows the spectrum of the Regge
pole also for the first configuration with the same parameters but for two different
high frequencies $2M_\infty\omega = 16$ and $32$. We can see that the structure
of the spectrum remains the same, however for very high frequencies (lower panel),
we can see the emergence of a new branch (purple triangles) between the two
branches associated with the inner and outer light-rings.

Fig.\ \ref{fig:PRs_2Mw_16_2Mw_32_Rs_5} shows the Regge pole spectrum
for $2M_\infty\omega =16$ and $32$ and this for the second configuration
($R_s >3M_\infty$) with parameters $M_\infty=1.5 M_{\text{\tiny BH}}$ and
$R_s = 5 M_{\text{\tiny BH}}$.  The structure remains the same for both frequencies
but the number of poles in the middle branch has increased with frequency.

The Regge poles fall into three distinct classes. We can see that they are relatively
similar to the Regge pole spectrum studied in the case of compact objects (see
\cite{OuldElHadj:2019kji}). Therefore, we will adopt the terminology already
introduced by these authors (see, also, Nussenzveig~\cite{nussenzveig2006diffraction}
for the origin of this terminology). We have thus :
\begin{enumerate}
\item \emph{Broad resonances}: poles with a relatively constant imaginary part which
is therefore parallel to the real axis with an approximately uniform spacing. They are
sensitive to the position of the shell and to its internal structure ($r < R_s$), i.e., its matter
content.

\item \emph{Inner surface waves}: strongly damped modes and their behaviour
are also entirely determined by the geometry of the object. The lowest modes
can be associated with waves propagating on the light-ring $r_{\ell_-}
= 3M_\text{\tiny BH}$, i.e., with the impact parameter \mbox{$b_{\ell_-}
=3\sqrt{3}M_{\textit{\tiny BH}}/\sqrt{\alpha}$}.

\item \emph{ Outer surface waves}: modes that depend essentially on the geometry
of the object. They are highly damped and the lowest modes can be associated with:
\begin{enumerate}[label=(\roman*)]
\item the \textit{surface waves} propagating on the outer light ring $r_{\ell_+} = 3M_\infty$,
i.e.\ with the impact parameter $b_{\ell_+} =3\sqrt{3} M_\infty$ for the configuration
where the shell is located between the two light-rings ($3M_\text{BH}<  R_s< 3M_\infty$).
\item the \textit{creeping modes} propagating along the shell surface at $r=R_s$, with impact
parameter $b_{R_S} = \sqrt{\frac{R_s^3}{R_s-2M_\infty}}$ for the configuration where
the shell is outside $3M_\infty$. They are generated by the edge rays (or grazing rays)
in the edge region.
\end{enumerate}
\end{enumerate}

In Figs~\ref{fig:PRs_2Mw_3_2Mw_6_Rs_4}, \ref{fig:PRs_2Mw_16_2Mw_32_Rs_4}
and \ref{fig:PRs_2Mw_16_2Mw_32_Rs_5} the outer surface waves (surface waves
and/or creeping modes), broad resonances, and inner surface waves are shown as
red squares, purple triangles, and blue circles, respectively. The black diamonds
represent the Regge poles of the isolated Schwarzschild black hole of mass
$M_{\text{\tiny BH}}$.

The lowest Regge poles are listed in (i) Table~\ref{tab:table2} for DBH with parameters
$M_\infty=1.5 M_{\text{\tiny BH}}$ and $R_s = 4 M_{\text{\tiny BH}}$ and for
$2M_\infty\omega = 3$, $6$, $16$ and $32$, and (ii) Table~\ref{tab:table3} for DBH
with parameters $M_\infty=1.5 M_{\text{\tiny BH}}$ and $R_s = 5 M_{\text{\tiny BH}}$
and for $2M_\infty\omega = 16$ and $32$.

\begingroup
%\squeezetable
\begin{table*}[htp]
\begin{threeparttable}[htp]
%\captionsetup{font=small}
\caption{\label{tab:table2} The lowest Regge poles $\lambda_{n}(\omega)$ for the
scalar field. The radius of the thin shell is $R_s = 4M_{BH}$ and the ADM mass is
$M_\infty = 1.5 M_{BH}$.}
\smallskip
\centering
\begin{ruledtabular}
\begin{tabular}{ccccc}
$2M_\infty\omega$  & $n$ & $\lambda^{\text{(O-S-W)\tnote{1}}}_n(\omega)$ &
$\lambda^{\text{(I-S-W)\tnote{2}}}_n(\omega)$ &
$\lambda^{\text{(B-R)\tnote{3}}}_n(\omega)$
\\ \hline
$3$  & $1$  & $7.539449 + 0.223849 i$  & $7.095909 + 1.687068 i $& $/ $ \\[-1ex]
& $2$  & $7.134677 + 0.790874 i$  & $7.022852 + 3.3708566 i$  & $/ $ \\[-1ex]
& $3$  & $6.956727 + 2.438228 i$  & $6.980154 + 5.790685 i $ & $/ $  \\[-1ex]
& $4$  & $6.934368 + 4.114637 i$  & $7.049810 + 7.549115 i $  & $/ $ \\[-1ex]
& $5$  & $7.154584 + 5.016518 i$  & $7.063185 + 9.371734 i $ & $/ $  \\[-1ex]
& $6$  & $7.390189 + 6.5269387i$  & $7.066195 + 11.160583 i $  & $/ $ \\[-1ex]
& $7$  & $7.652774 + 7.865429 i$  & $7.067851 + 12.940759 i$ & $/ $  \\[-1ex]
& $8$  & $8.001365 + 9.087813 i$  & $7.062229 + 14.719719 i $ & $/ $  \\[-1ex]
& $9$  & $8.362859 + 10.289717i$  & $7.051567 + 16.502463 i $  & $/ $ \\[-1ex]
& $10$ & $8.715833 + 11.441810i$  & $7.039533 + 18.290428 i $   & $/ $   \\[+1ex]
$6$   & $1$  & $15.370841+0.179585 i$  & $14.611707+1.414356 i  $ & $/ $  \\[-1ex]
& $2$  & $14.548787+0.528288 i $  & $14.209917+3.071463 i$  & $/ $ \\[-1ex]
& $3$  & $14.255707+1.927388 i $  & $13.820181+4.931434 i $ & $/ $  \\[-1ex]
& $4$  & $14.075994 + 3.476095 i $  & $13.520911+6.640027 i $ & $/ $  \\[-1ex]
& $5$  & $14.169925+4.925956 i $  & $13.303758+8.406231 i  $ & $/ $  \\[-1ex]
& $6$  & $14.352709+6.542016 i $  & $13.132015+10.207303 i$  & $/ $ \\[-1ex]
& $7$  & $14.572460+8.082838 i $  & $12.995979+12.028643 i $  & $/ $ \\[-1ex]
& $8$  & $14.832787+9.558134 i $  & $12.889802+13.859215 i $  & $/ $ \\[-1ex]
& $9$  & $15.124109+10.975998 i $  & $12.808015+15.691713 i $  & $/ $ \\[-1ex]
& $10$  & $15.437413+12.345040 i $  & $12.745238+17.521595 i $ & $/ $ \\[+1ex]
$16$ & $1$ & $41.670935 + 0.046352 i $ & $39.308955 + 0.526143 i$& $/ $ \\[-1ex]
& $2$ & $40.757178 + 0.815017 i $ & $39.123982 + 0.945890 i $ & $/ $\\[-1ex]
& $3$ & $39.944801 + 2.350849 i$ & $38.477482 + 2.002715 i$& $/ $ \\[-1ex]
& $4$ & $39.562328 + 4.058189 i $ & $37.936546 + 3.242729 i$& $/ $ \\[-1ex]
& $5$ & $39.354113 + 5.746829 i$ & $37.428640 + 4.601256 i $ & $/ $\\[-1ex]
& $6$ & $39.253632 + 7.417623 i $ & $36.946769 + 6.047137 i $& $/ $ \\[-1ex]
& $7$ & $39.231469 + 9.068162 i $ & $36.487997 + 7.562487 i $ & $/ $\\[-1ex]
& $8$ & $39.269947 + 10.696820 i$ & $36.051294 + 9.134979 i $ & $/ $\\[-1ex]
& $9$ & $39.357086 + 12.302735 i$ & $35.636215 + 10.755451 i $& $/ $ \\[-1ex]
& $10$ & $39.484150 + 13.885587 i $ & $35.242599 + 12.416733 i $ & $/ $\\[+1ex]
$32$ & $1$   & $83.931483+0.001076 i $  & $78.343315+0.569228 i$   & $78.717561+0.863328 i$ \\[-1ex]
& $2$ & $82.718881+0.302994 i $  & $77.708251+1.655406 i $  & $80.826502+1.279706 i$\\[-1ex]
& $3$ & $81.765639+1.257533 i $  & $77.049161+2.705208 i $  & $ / $\\[-1ex]
& $4$ & $80.934661+2.942484 i $  & $76.407917+3.858777 i $  & $ / $\\[-1ex]
& $5$ & $80.484718+4.574505 i $  & $75.788245+5.093969 i $  & $ /$\\[-1ex]
& $6$ & $80.144197+6.211451 i $  & $75.189451+6.395209 i $  & $/ $ \\[-1ex]
& $7$ & $79.887645+7.854207 i $  & $74.609971+7.751822 i $  & $/ $ \\[-1ex]
& $8$ & $79.697425+9.498636 i $  & $74.048266+9.156061 i $  & $/ $ \\[-1ex]
& $9$ & $79.561618+11.141534 i$  & $73.503036+10.602050 i $  & $/ $ \\[-1ex]
& $10$ &$79.471725+12.780560 i $  & $72.973244+12.085172 i $  & $/ $ \\
\end{tabular}
\end{ruledtabular}
\begin{tablenotes}
\item[1] O-S-W : Outer surface waves
\item[2] I-S-W : Inner surface waves
\item[3] B-R   : Broad resonances
\end{tablenotes}
\end{threeparttable}
\end{table*}
\endgroup

\begingroup
%\squeezetable
\begin{table*}[htp]
%\fontsize{5.}{10}
\begin{threeparttable}[htp]
%\captionsetup{font=small}
\caption{\label{tab:table3} The lowest Regge poles $\lambda_{n}(\omega)$ for the
scalar field. The radius of the thin shell is $R_s = 5M_{BH}$ and the ADM mass is
$M_\infty = 1.5M_{BH}$.}
\smallskip
\centering
\begin{ruledtabular}
\begin{tabular}{ccccc}
$2M_\infty\omega$  &$n$  & $\lambda^{\text{(O-S-W)\tnote{1}}}_n(\omega)$  &
$\lambda^{\text{(I-S-W)\tnote{2}}}_n(\omega)$ &
$\lambda^{\text{(B-R)\tnote{3}}}_n(\omega)$
\\ \hline
$16$  & $1$  & $41.860118 + 1.784810 i  $  & $33.945579 + 0.477813 i $
& $35.086529 + 1.294345 i$   \\[-1ex]
& $2$  & $42.120571 + 3.964122 i  $  & $33.695542 + 1.278693 i $
& $37.393437 + 1.734247 i$   \\[-1ex]
& $3$  & $42.451238 + 6.041693 i  $  & $33.057993 + 2.190304 i $  & $/ $   \\[-1ex]
& $4$  & $42.818104 + 8.030389 i  $  & $32.399192 + 3.251233 i $  & $/ $   \\[-1ex]
& $5$  & $43.205922 + 9.949368 i  $  & $31.759081 + 4.409909 i $  & $/ $   \\[-1ex]
& $6$  & $43.608414 + 11.811534 i  $  & $31.143492 + 5.647996 i $  & $/ $   \\[-1ex]
& $7$  & $44.022224 + 13.625389 i  $  & $30.554809 + 6.952932 i $  & $ /$   \\[-1ex]
& $8$  & $44.445123 + 15.396893 i  $  & $29.994263 + 8.315144 i $  & $/ $   \\[-1ex]
& $9$  & $44.875425 + 17.130499 i  $  & $29.462624 + 9.726841 i $  & $/ $   \\[-1ex]
& $10$  & $45.311767 + 18.829705 i  $  & $28.960402 + 11.181362 i $  & $/ $  \\[+1ex]
$32$ & $1$ & $84.123387 + 2.204783 i $  & $67.582661 + 1.284484 i$
& $68.724166 + 1.194667 i $\\[-1ex]
& $2$ & $84.576124 + 4.572500 i $  & $66.893567 + 2.072605 i$
& $70.375698 + 1.488244 i $\\[-1ex]
& $3$ & $85.005575 + 6.802744 i$   & $66.166349 + 2.972146 i$
& $72.419573 + 1.755117 i $\\[-1ex]
& $4$ & $85.428874 + 8.940819 i $  & $65.438785 + 3.951276 i$
& $74.941664 + 2.039113 i $\\[-1ex]
& $5$ & $85.849013 + 11.011312 i$  & $64.718695 + 4.996764 i$
& $78.206962 + 2.394064 i $\\[-1ex]
& $6$ & $86.267934 + 13.029013 i$  & $64.009870 + 6.099179 i$ & $/ $ \\[-1ex]
& $7$ & $86.687058 + 15.003322 i$  & $63.314252 + 7.251465 i$ & $/ $ \\[-1ex]
& $8$ & $87.107350 + 16.940586 i$  & $62.632891 + 8.448162 i$ & $/ $ \\[-1ex]
& $9$ & $87.529435 + 18.845339 i$  & $61.966385 + 9.684914 i$ & $/ $ \\[-1ex]
& $10$ &$87.953689 + 20.720964 i$  & $61.315094 + 10.958141i$ & $/ $ \\
\end{tabular}
\end{ruledtabular}
\begin{tablenotes}
\item[1] O-S-W : Outer surface waves (creeping modes)
\item[2] I-S-W : Inner surface waves
\item[3] B-R   : Broad resonances
\end{tablenotes}
\end{threeparttable}
\end{table*}
\endgroup

\subsection{The WKB approximation}

In order to gain insight into the physical origin of the broad resonances identified
in Figs.~\ref{fig:rp1} and~\ref{fig:rp2}, we now turn to the WKB approximation.

\subsubsection{Propagation of WKB modes}

We begin by constructing the WKB solution to \eqref{H_Radial_equation}
by assuming that solutions are rapidly oscillating functions with slowly varying amplitudes:
\begin{equation}
\phi_\ell (r_*) = A(r_*)e^{i \int p(r'_*) dr_*'}.
\end{equation}

Inserting this ansatz into the wave equation, and assuming that $|p'| \ll |p^2| $
and $|A'| \ll |pA|$, we get that
\begin{equation}
p(r_*)^2 + V(r_*) = 0, \quad A = a |p|^{-1/2},
\end{equation}
where $a$ is a constant coefficient, $V = V_\ell - \omega^2$, and we suppress
$\omega$ and $\ell$ indices for clarity. The first of the above equations is quadratic
in $p$, thus it will admit two solutions that can be interpreted as waves travelling radially
inward and outward. Taking into account contributions from both modes, we can write
the following WKB solution to the wave equation:
\begin{equation}
\phi_\ell = |p|^{-1/2}\left( a^{in} e^{-i\int{p dr_*}} + a^{out} e^{i\int{p dr_*}}\right).
\end{equation}
We see from the above expression that, in regions where the WKB approximation holds,
the only change to the amplitudes of the modes comes from the variation of the function $p$.
It is also clear that the WKB approximation breaks down at those points which satisfy
$V(r_*) = 0$, since it implies that $p = 0$, leading to an infinite amplitude of the WKB modes.
The points satisfying $V(r_*) = 0$, are called turning points and one can write a WKB
solution on both sides of the turning points, with different coefficients
$(a^{in},a^{out})$ on each side.
One then relates the WKB modal coefficients on each side of the turning points by
applying connection formulae (see~\cite{Berry_1972}).
Between turning points, the WKB modes do not mix and their propagation between
two points $r_{*i}$ and $r_{*j}$ is captured by the following propagation matrices:
\begin{equation}
\binom{a^{out}_i}{a^{in}_i} = P_{ij}\binom{a^{out}_j}{a^{in}_j},
\end{equation}
where
\begin{equation}
P_{ij} = \begin{pmatrix}
e^{-iS_{ij}} & 0 \\
0 & e^{iS_{ij}}
\end{pmatrix} \quad \text{when}\quad  p^2(r_{*i}<r_*<r_{*j}) > 0,
\end{equation}
or
\begin{equation}
P_{ij} = \begin{pmatrix}
0         & e^{-S_{ij}} \\
e^{S_{ij}} &          0
\end{pmatrix} \quad \text{when}\quad  p^2(r_{*i}<r_*<r_{*j}) < 0.
\end{equation}
In the above formula, $S_{ij}$ is the WKB action between the points $r_{*i}$ and $r_{*j}$
and is given by
\begin{equation}
S_{ij} = \int_{r_{*i}}^{r_{*j}} |p(r_*)|dr_*.
\end{equation}
Combining the propagation matrix for modes below the potential barrier and the
connection matrices across isolated turning points, we define
the tunneling matrix, $T_{12}$, which connects WKB modes on the outside of
two turning points $r_{*_1}$ and $r_{*_2}$:
\begin{equation}
T_{12} = \begin{pmatrix}
1/\mathcal{T} & -\mathcal{R}/\mathcal{T}, \\
\mathcal{R}^*/\mathcal{T} & 1/\mathcal{T}^*.
\end{pmatrix},
\end{equation}
where $\mathcal{R}$ and $\mathcal{T}$ are the local reflection and transmission
coefficients across the potential barrier and are given by:
\begin{eqnarray}
\mathcal{R} &=& -i\frac{1-e^{-2S_{12}}/4}{1+e^{-2S_{12}}/4}\\
\mathcal{T} &=& \frac{e^{-2S_{12}}}{1+e^{-2S_{12}}/4}
\end{eqnarray}

\subsubsection{WKB modes across the shell}

In order to construct the full WKB solution across the entire radial range, we also
need to connect WKB modes on both sides of the shell. Since we are only interested
in the qualitative behaviour of the WKB modes in order to interpret the broad
resonance, we assume here that the jump discontinuity in the potential is much
smaller than the frequencies of the WKB modes, $\omega$, and that this frequency
does not vary significantly near the shell.
Within this framework, we write the WKB solution for $r\sim R_s$ as:
\begin{equation}
\phi_{\omega,l}(r_*) = \begin{cases} a^{out}_{L}e^{i\omega r_*}
+ a^{in}_L e^{-i\omega r_*} \ \text{for} \ r_* \leq R_s^*, \\
a^{out}_{R}e^{i\omega r_*} + a^{in}_R e^{-i\omega r_*} \ \text{for} \ r_* \geq R_s^*.
\end{cases}
\end{equation}
From the junction condition at the shell given in \eqref{jump_condition}, we can
define the connection matrix, $\mathcal{C}$, which connects the WKB modal
coefficients on the left and right sides of the shell:
\begin{equation}
\mathcal{C} = \begin{pmatrix}
1 & - e^{-2i\omega R_s} \frac{\Delta}{2i\omega} \\
e^{2i\omega R_s} \frac{\Delta}{2i\omega} & 1
\end{pmatrix},
\end{equation}
where $\Delta = \frac{\sqrt{AB}_+ - \sqrt{AB}_{-}}{R_s}$. Note that $\Delta$ is
not the discontinuity of the potential but rather the coefficient entering in the discontinuity
of the field's derivative. From the connection matrix, we can also define a reflection
coefficient across the shell:
\begin{eqnarray}
\mathcal{R}_s = \frac{\Delta}{2i\omega}.
\end{eqnarray}
Note here that the reflection across the discontinuity is of order $\mathcal{O}(\omega^{-1})$.
This is due to the fact that the metric itself is discontinuous at the shell and is in contrast
with the case of a compact object which usually has a continuous metric and for which
the reflection coefficient at the surface of the object is at least of order
$\mathcal{O}(\omega^{-2})$~\cite{Berry_1982,Zhang:2011pq}.

\subsubsection{WKB estimate of the broad resonances}

We now have all the necessary quantities to relate the modal coefficient at
$r_{*}\rightarrow -\infty$ to the one at $r_* \rightarrow + \infty$.
This is done by combining the propagating, tunneling and connection matrix as
follows (see Fig.\ \ref{fig:WKB} for illustration):

\begin{figure}
\includegraphics[trim=1cm 0 0 0, scale=0.7]{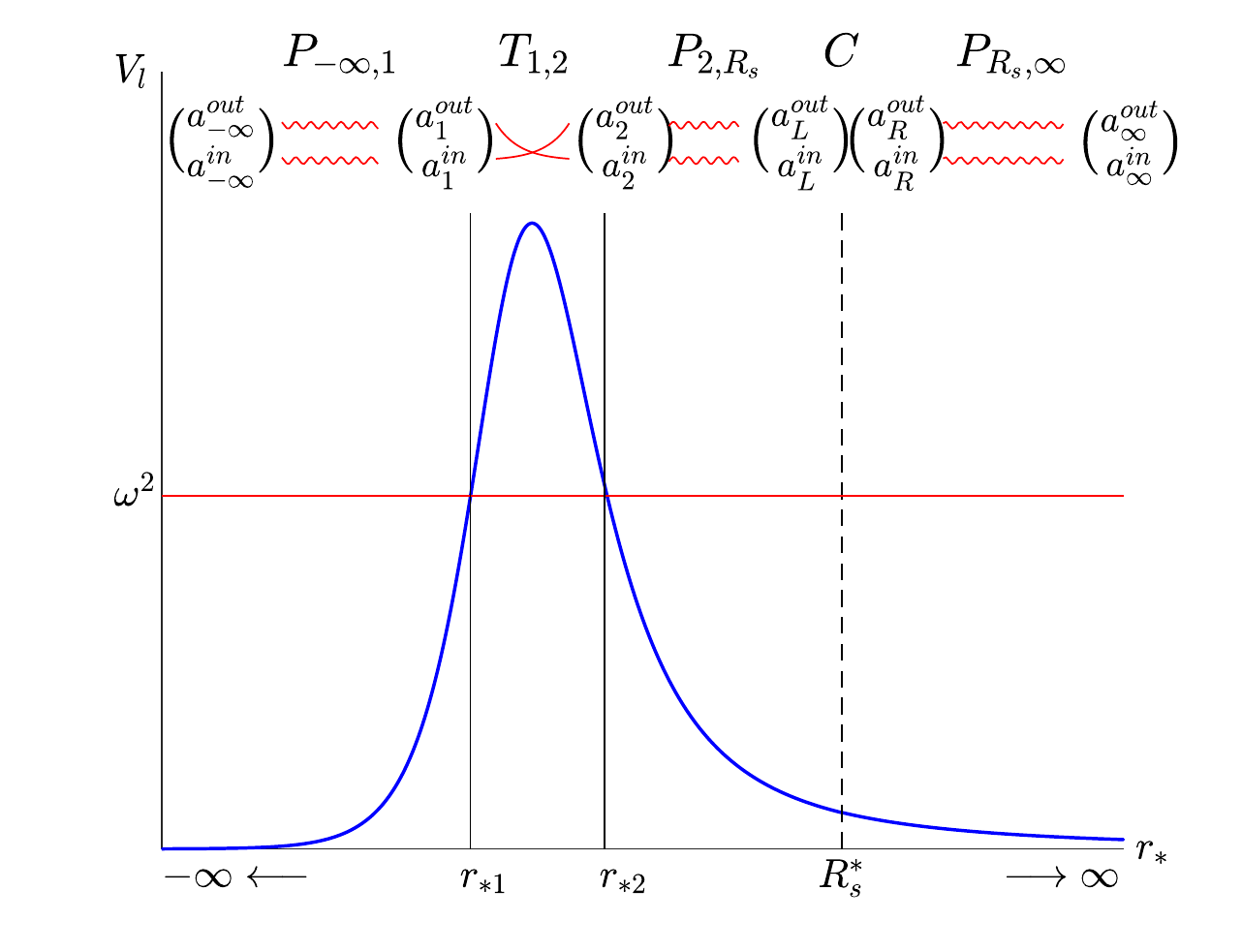}
\caption{Illustration of the connection formula \eqref{eq:connection_WKB}
obtained by combining the propagating, tunneling and connection matrix to relate
the WKB modal coefficient across the turning points and the shell. The oscillating
(growing/decaying) lines represent the propagation of WKB modes with real
(imaginary) momenta $p$. }
\label{fig:WKB}
\end{figure}

\begin{equation}\label{eq:connection_WKB}
\binom{a^{out}_{-\infty}}{a^{in}_{-\infty}} = P_{-\infty 1}\times T_{12} \times P_{2R_s}
\times \mathcal{C} \times P_{R_s\infty} \binom{a^{out}_{\infty}}{a^{in}_\infty}.
\end{equation}

Solving the above system for outgoing boundary conditions, that is $a^{out}_{-\infty}
= a^{in}_{\infty}=0$, leads to the following condition:
\begin{equation}
e^{-2iS_{2R_s} - 2i\omega R_s} = \mathcal{R}\mathcal{R}_s.
\end{equation}

We now anticipate that broad resonances will be slowly damped and therefore the
real part of their RPs will be much greater than their imaginary part. We write
$\lambda = \lambda_R + i\Gamma$, with $\Gamma \ll \lambda_r$.
The real and imaginary parts of the above condition give
\begin{equation}
\sin\left( S_{2R_S}(\lambda_R) + \omega R_s - \frac{\pi}{4} \right) = 0 \ \ \text{and}
\ \ \Gamma = \frac{\ln(|\mathcal{R}\mathcal{R}_s|)}{2\partial_\lambda S_{2Rs}
\big|_{\lambda_R}}.
\end{equation}
Hence the location of the shell sets the real part of the broad resonances part of the
RP spectrum while its matter content sets their imaginary part. Contrary to the other
branches of the RP spectrum, the above systems admit a finite number of solutions.
In particular we must have that the turning points be located between the outer
light-ring and the shell, which limits the range of $\lambda_R$.

Solving the above condition numerically for the shell parameters of Fig.\ \ref{fig:rp1},
we find that there are only two solutions, $\lambda_{\mathrm{WKB}}^{(1)}
= 35.6946 + 1.79675 i $ and $\lambda_{\mathrm{WKB}}^{(2)} = 38.2285 + 2.8487 i$,
which differ from the values found numerically by $2.2\%$ and $3.7\%$ respectively.

\section{Wave Scattering by a dirty black hole}\label{sec:scattering}

In this section, we compute the differential scattering cross sections $d\sigma/d\Omega$
for plane monochromatic scalar waves impinging upon a DBH using the partial wave
expansion, and we compare with the results constructed by CAM representations of
these cross sections by means of the Sommerfeld-Watson transform and Cauchy
theorem~\cite{Newton:1982qc,Watson18,Sommerfeld49}.

\subsection{The differential scattering cross section: Partial waves expansion}

The differential scattering cross section for a scalar field is given by \cite{Futterman:1988ni}
\begin{equation}\label{Scalar_Scattering_diff}
\frac{d\sigma}{d\Omega} = |f(\omega,\theta)|^2
\end{equation}
where
\begin{equation}\label{Scalar_Scattering_amp}
f(\omega,\theta) = \frac{1}{2 i \omega} \sum_{\ell = 0}^{\infty} (2\ell+1)
[S_{\ell}(\omega)-1]P_{\ell}(\cos\theta)
\end{equation}
denotes the scattering amplitude. In \eqref{Scalar_Scattering_amp}, the
functions $P_{\ell}(\cos\theta)$ are the Legendre polynomials~\cite{AS65} and
the $S$-matrix elements $S_{\ell}(\omega)$ were given by \eqref{Matrix_S}.

\subsection{CAM representation of the scattering amplitude}

Following the steps in \cite{Folacci:2019cmc}, we construct the CAM
representation of $f(\theta)$ using a Sommerfeld-Watson transformation
\cite{Watson18,Sommerfeld49,Newton:1982qc}
\begin{equation}
\label{SWT_gen}
\sum_{\ell=0}^{+\infty} (-1)^\ell F(\ell)= \frac{i}{2} \int_{\cal C} d\lambda \,
\frac{F(\lambda -1/2)}{\cos (\pi \lambda)} ,
\end{equation}
where $F(\cdot)$ is any function without singularities on the real axis $\lambda$.
By means of \eqref{SWT_gen}, we replace the discrete sum over the
ordinary angular momentum $\ell$ in \eqref{Scalar_Scattering_amp} with a
contour integral in the complex $\lambda$ plane (i.e., in the complex
$\ell$-plane with $\lambda = \ell +1/2$). By noting that $P_\ell (\cos \theta)
=(-1)^\ell P_\ell (-\cos \theta)$, we obtain
\begin{eqnarray}
\label{SW_Scalar_Scattering_amp}
& & f(\omega,\theta) = \frac{1}{2 \omega}  \int_{\cal C} d\lambda \, \frac{\lambda}
{\cos (\pi \lambda)} \nonumber \\
&&  \qquad\qquad   \times \left[ S_{\lambda -1/2} (\omega) -1 \right]P_{\lambda -1/2}
(-\cos \theta).
\end{eqnarray}
It should be noted that, in \eqref{SWT_gen} and \eqref{SW_Scalar_Scattering_amp},
the integration contour encircles counterclockwise the positive real axis of the complex
$\lambda$-plane, and $P_{\lambda -1/2} (z)$
denotes the analytic extension of the Legendre polynomials $P_\ell (z)$ which is
defined in terms of hypergeometric functions by~\cite{AS65}
\begin{equation}\label{Def_ext_LegendreP}
P_{\lambda -1/2} (z) = F[1/2-\lambda,1/2+\lambda;1;(1-z)/2].
\end{equation}
Here, $S_{\lambda -1/2} (\omega)$ is given by [see \eqref{Matrix_S}]
\begin{equation}\label{Matrix_S_CAM}
S_{\lambda -1/2}(\omega) =  e^{i(\lambda + 1/2)\pi} \,
\frac{A_{\lambda -1/2}^{(+)}(\omega)}{A_{\lambda -1/2}^{(-)}(\omega)}
\end{equation}
and denotes ``the'' analytic extension of $S_\ell (\omega)$ where the complex
amplitudes $A^{(\pm)}_{\lambda -1/2} (\omega)$ are defined from the analytic
extension of the modes $\phi_{\omega \ell}$, i.e., from the function
$\phi_{\omega ,\lambda -1/2}$.

It is also important to recall that the Regge poles $\lambda_n(\omega)$ of
$S_{\lambda-1/2}(\omega)$ lie in the first and third quadrants,
symmetrically distributed with respect to the origin $O$, and are defined as the
zeros of the coefficient  $A^{(-)}_{\lambda-1/2} (\omega)$ [see \eqref{Matrix_S_CAM}]
\begin{equation}\label{PR_def_Am}
A^{(-)}_{\lambda_n(\omega)-1/2} (\omega)=0,
\end{equation}
with $n=1,2,3,\ldots$., and the associated residues at the poles
$\lambda=\lambda_n(\omega)$ are defined by [see \eqref{Matrix_S_CAM}]
\begin{equation}\label{residues_RP}
r_n(\omega)=e^{i\pi [\lambda_n(\omega)+1/2]} \left[ \frac{A_{\lambda -1/2}^{(+)}
(\omega)}{\frac{d}{d \lambda}A_{\lambda -1/2}^{(-)}(\omega)}
\right]_{\lambda=\lambda_n(\omega)}.
\end{equation}

In order to collect the Regge pole contributions, we deform the contour ${\cal C}$
in \eqref{SW_Scalar_Scattering_amp} while using the Cauchy theorem to obtain
\begin{equation}\label{CAM_Scalar_Scattering_amp_tot}
f (\omega, \theta) =  f^\text{\tiny{B}} (\omega, \theta) +  f^\text{\tiny{RP}} (\omega, \theta)
\end{equation}
where
\begin{subequations}\label{CAM_Scalar_Scattering_amp_decomp}
\begin{equation}\label{CAM_Scalar_Scattering_amp_decomp_Background}
f^\text{\tiny{B}} (\omega, \theta) = f^\text{\tiny{B},\tiny{Re}} (\omega, \theta)
+ f^\text{\tiny{B},\tiny{Im}} (\omega, \theta)
\end{equation}
is a background integral contribution with
\begin{equation}\label{CAM_Scalar_Scattering_amp_decomp_Background_a}
f^\text{\tiny{B},\tiny{Re}} (\omega, \theta) = \frac{1}{\pi \omega} \int_{{\cal C}_{-}}
d\lambda \, \lambda S_{\lambda -1/2}(\omega) Q_{\lambda -1/2}(\cos \theta +i0)
\end{equation}
and
\begin{eqnarray}\label{CAM_Scalar_Scattering_amp_decomp_Background_b}
f && ^\text{\tiny{B},\tiny{Im}}  (\omega, \theta) = \frac{1}{2 \omega}\left(\int_{+i\infty}^{0}
d\lambda \, \left[S_{\lambda -1/2}(\omega) P_{\lambda-1/2} (-\cos \theta) \right. \right.
\nonumber \\
- && \left.  \left. S_{-\lambda -1/2}(\omega) e^{i \pi \left(\lambda+1/2\right)}P_{\lambda-1/2}
(\cos \theta) \right] \frac{\lambda}{\cos (\pi \lambda) } \right).
\end{eqnarray}
\end{subequations}
The second term in \eqref{CAM_Scalar_Scattering_amp_tot}
\begin{eqnarray}
\label{CAM_Scalar_Scattering_amp_decomp_RP}
& & f^\text{\tiny{RP}} (\omega, \theta) = -\frac{i \pi}{\omega}    \sum_{n=1}^{+\infty}   \frac{ \lambda_n(\omega) r_n(\omega)}{\cos[\pi \lambda_n(\omega)]}  \nonumber \\
&&  \qquad\qquad \qquad\qquad \times  P_{\lambda_n(\omega) -1/2} (-\cos \theta),
\end{eqnarray}
is a sum over the Regge poles lying in the first quadrant of the CAM plane.
Of course, the CAM representation of the scattering amplitude $f (\omega, \theta)$
for the scalar field given by \eqref{CAM_Scalar_Scattering_amp_tot} and
\eqref{CAM_Scalar_Scattering_amp_decomp} is equivalent to the initial partial wave
expansion \eqref{Scalar_Scattering_amp}. From this CAM representation, we
extract the contribution $f^\text{\tiny{RP}}(\omega, \theta)$ given by
\eqref{CAM_Scalar_Scattering_amp_decomp_RP} which is only an approximation
of $f(\omega, \theta)$, and which provides us with a corresponding approximation of
the differential scattering cross section via (\ref{Scalar_Scattering_diff}).

\subsection{Computational methods}

In order to construct the scattering amplitude \eqref{Scalar_Scattering_amp}, and
the Regge pole contribution \eqref{CAM_Scalar_Scattering_amp_decomp_RP},
it is necessary first to obtain the function $\phi_{\omega\ell}^{\text{\tiny in}}(r)$,
the coefficients $A_\ell^{(\pm)}(\omega)$, and the $S$-matrix elements
$S_\ell(\omega)$ by solving \eqref{H_Radial_equation} with conditions \eqref{bc_in},
and second to compute the Regge poles $\lambda_n(\omega)$ of \eqref{PR_def_Am} and
the associated residues \eqref{residues_RP}. To do this, we use the numerical methods
of \cite{Folacci:2019cmc,Folacci:2019vtt} (see Secs.\ III B and IVA of these papers).
It is important to note that, the scattering amplitude \eqref{Scalar_Scattering_amp} suffers
a lack of convergence due to the long range nature of the field propagating on the
Schwarzschild spacetime (outside the thin shell) and to accelerate the convergence
of this sum, we have used the method described in the Appendix of
\cite{Folacci:2019cmc}. All numerical calculations were performed using {\it Mathematica}.

\subsection{Numerical Results and comments: Scattering cross sections}

\begin{figure*}[htp!]
\includegraphics[scale=0.50]{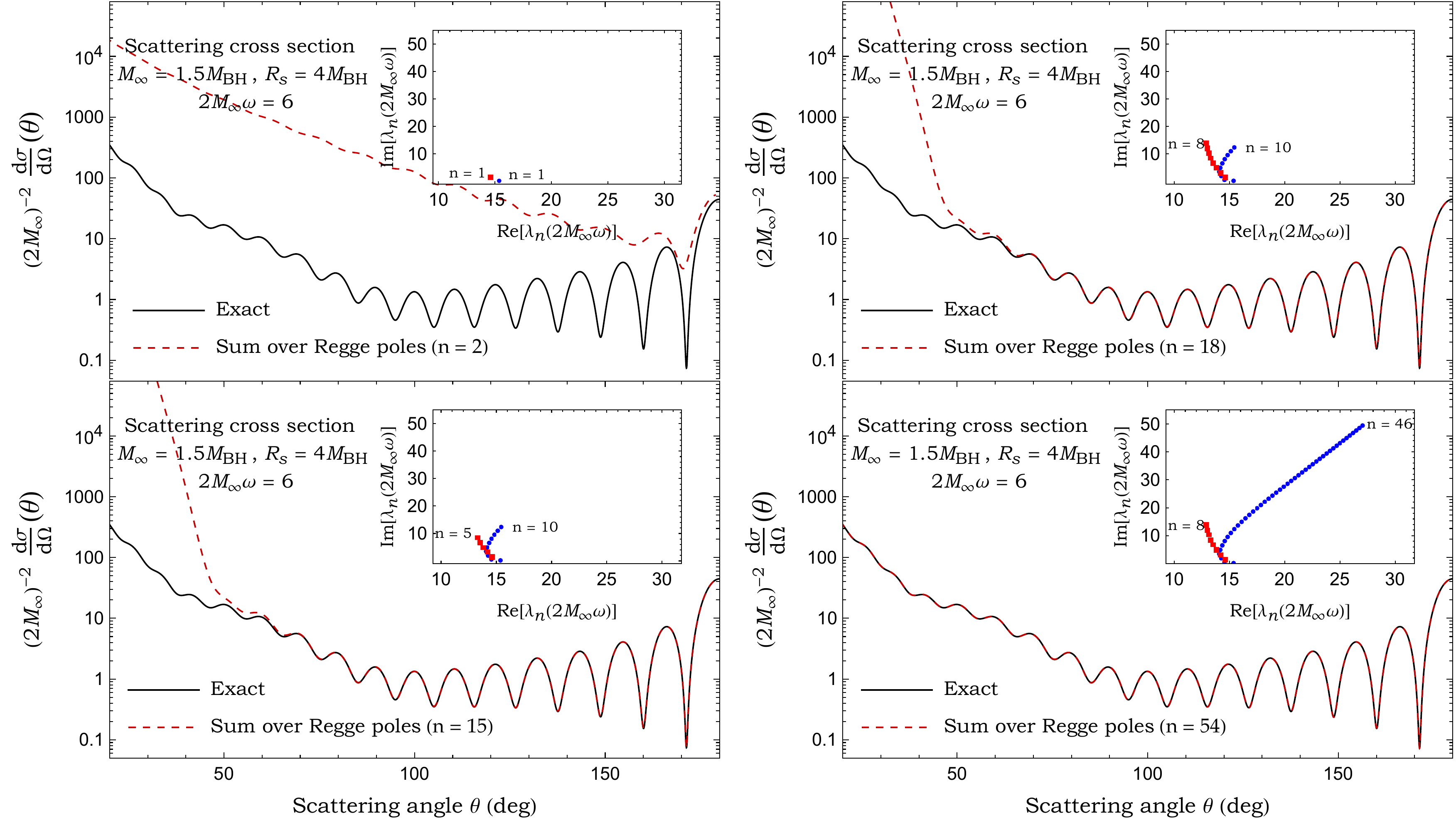}
%\vspace*{-0.35cm}
\caption{\label{fig:CAM_Sec_2Mw_6_PRs_Rs_4} The scalar cross section of
DBH for $2M_\infty\omega=6$,
$R_s=4M_{\text{\tiny BH}} $ and $M_\infty=1.5M_{\text{\tiny BH}} $ and its
Regge pole approximation. The plots show the effect of including successively
more Regge poles.}
\end{figure*}

\begin{figure*}[htp!]
\includegraphics[scale=0.50]{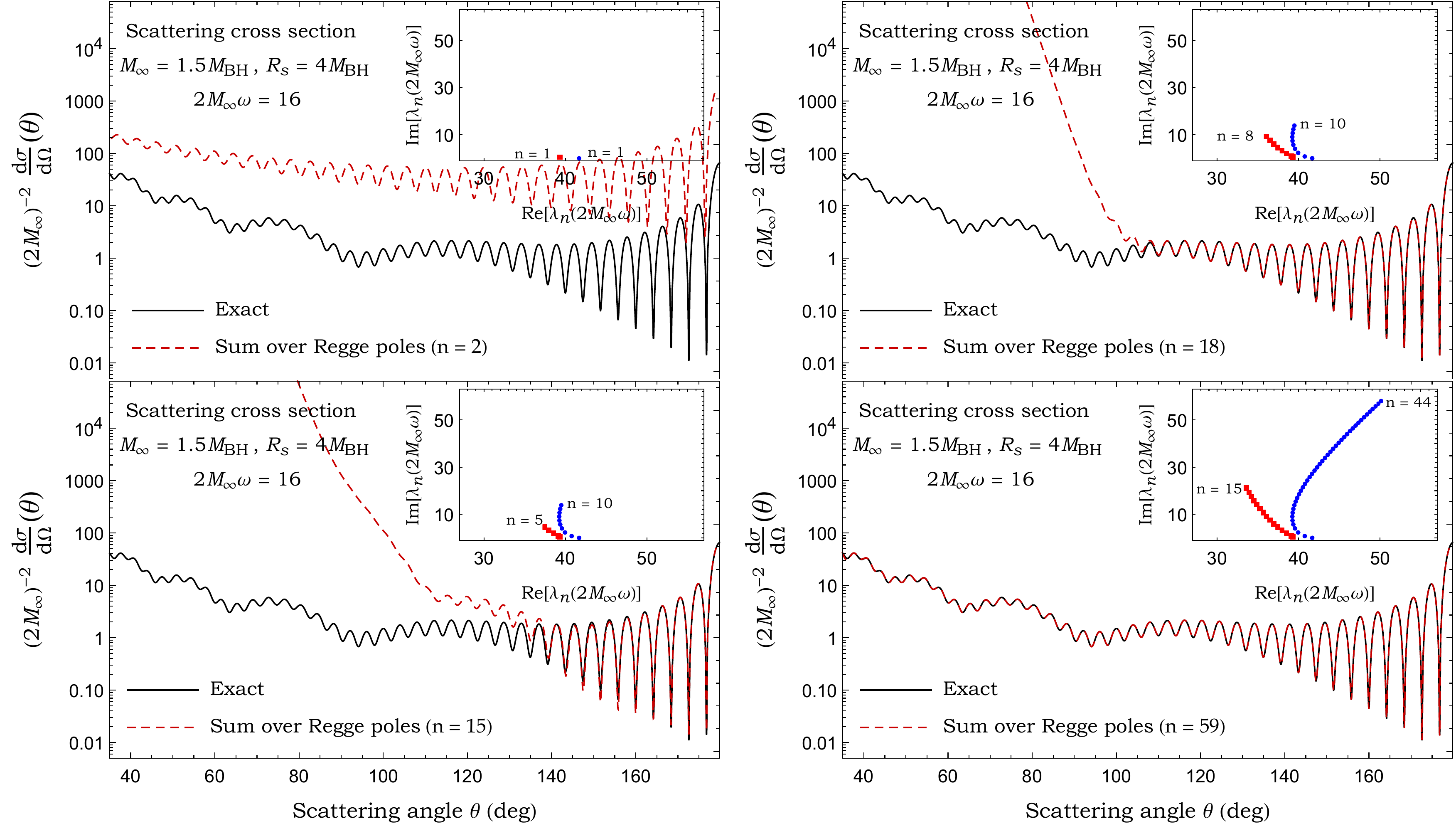}
%\vspace*{-0.35cm}
\caption{\label{fig:CAM_Sec_2Mw_16_PRs_Rs_4} The scalar cross section of DBH
for $2M_\infty\omega=16$, $R_s=4M_{\text{\tiny BH}}$ and $M_\infty=1.5M_{\text{\tiny BH}}$
and its Regge pole approximation. The plots show the effect of including successively
more Regge poles.}
\end{figure*}

\begin{figure*}[htp!]
\includegraphics[scale=0.50]{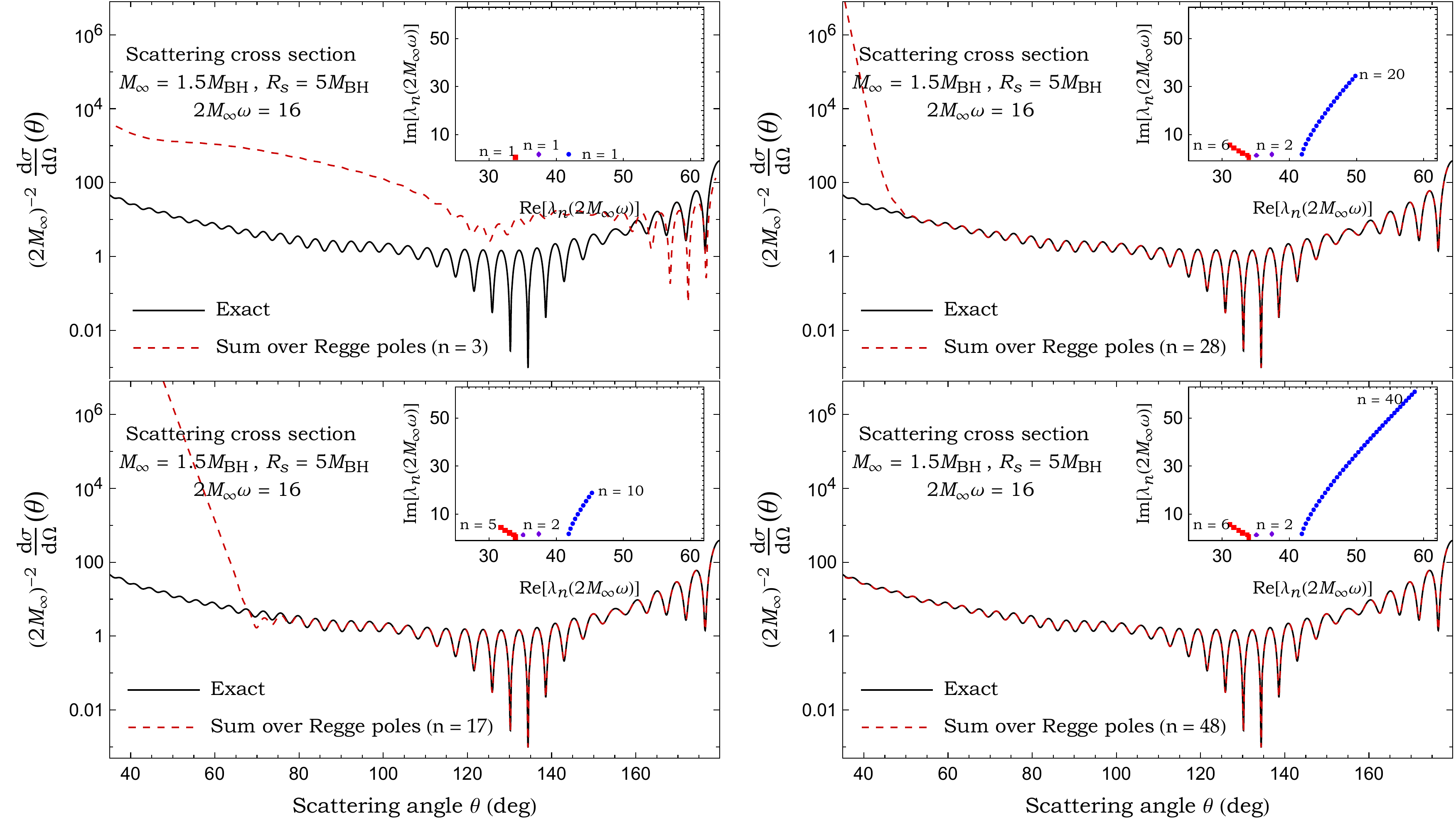}
%\vspace*{-0.35cm}
\caption{\label{fig:CAM_Sec_2Mw_16_Rs_5_PRs}  The scalar cross section of
DBH for $2M_\infty\omega=16$, $R_s=5M_{\text{\tiny BH}} $ and
$M_\infty=1.5M_{\text{\tiny BH}} $ and its Regge pole approximation.
The plots show the effect of including successively more Regge poles.}
\end{figure*}

\begin{figure}[htp!]
\includegraphics[scale=0.50]{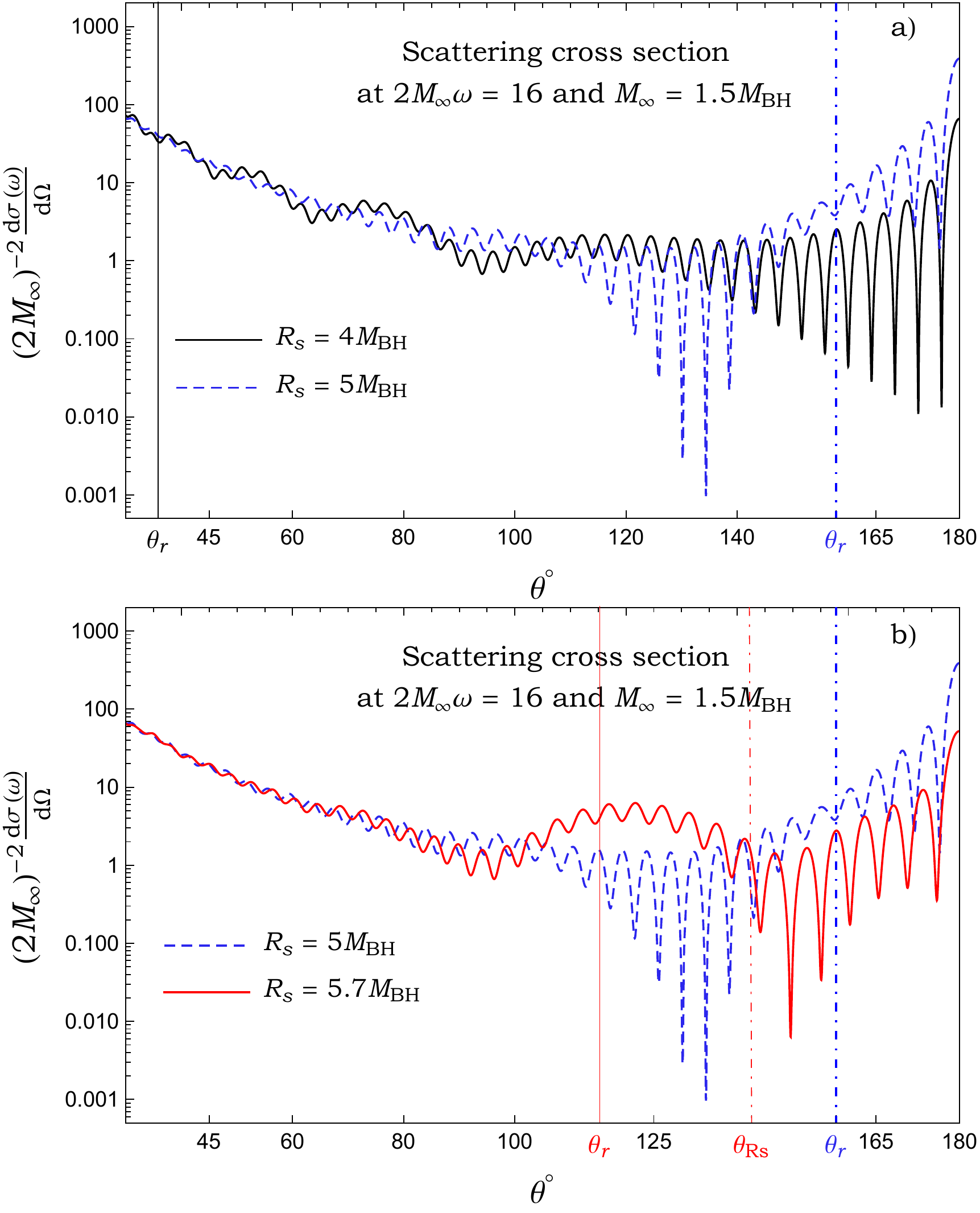}
%\vspace*{-0.35cm}
\caption{\label{fig:Sactt_Cross_section_Rs-4_Vs_Rs-5_2_5dot7_2Mw_16}
The scalar cross section of
DBH for $2M_\infty\omega=16$,  and $M_\infty=1.5M_{\text{\tiny BH}} $.
a) We compare $R_s=5M_{\text{\tiny BH}}$ to $R_s=4M_{\text{\tiny BH}}$.
We can see that the scattering amplitude is enhanced by the rainbow effect
for  $R_s=5M_{\text{\tiny BH}}$ at the rainbow angle $\theta_r\approx 157.8^{\circ}$
(blue dotdashed line). b) We compare $R_s=5M_{\text{\tiny BH}}$ to
$R_s=5.7M_{\text{\tiny BH}}$. We can see that the scattering amplitude is
enhanced at $\theta_r\approx 115.3^{\circ}$ (red line) for $R_s=5M_{\text{\tiny BH}}$
and at $\theta_r\approx 157.8^{\circ}$  (blue dotdashed line).
}\label{fig:comparison_critical}
\end{figure}

We present in Figs~\ref{fig:CAM_Sec_2Mw_6_PRs_Rs_4},
\ref{fig:CAM_Sec_2Mw_16_PRs_Rs_4} and \ref{fig:CAM_Sec_2Mw_16_Rs_5_PRs}
various scattering cross sections constructed from the CAM approach, and
compare with results obtained from the partial wave expansion method.
As in Sec.\ \ref{sec:resonances}, we focus on two configurations of the DBH:
(i) a DBH with  $3M_{\text{\tiny BH}} < R_s < 3M_\infty$, and
(ii) DBH with $R_s >3M_\infty$.

Fig.\ \ref{fig:CAM_Sec_2Mw_6_PRs_Rs_4} shows the scattering cross section
constructed from the CAM approach at $2M_\infty\omega =6$ for a shell configuration
with $M_\infty=1.5 M_{\text{\tiny BH}}$ and $R_s=4M_{\text{\tiny BH}}$, i.e.\ for a
DBH configuration with $3 M_{\text{\tiny BH}} < R_s< 3M_\infty$. We can see that it
gets progressively closer to the scattering cross section constructed from partial
wave expansion by including increasingly more Regge poles in the sum
\eqref{CAM_Scalar_Scattering_amp_decomp_RP}. Indeed, the sum over only the
first two Regge poles in equation~\eqref{CAM_Scalar_Scattering_amp_decomp_RP}
does not reproduce the cross section, however, summing over $54$ Regge poles, but
without including the background integral, the agreement is perfect with that constructed
from the partial wave expansion for $\theta \gtrsim 20^\circ $.

Fig.\ \ref{fig:CAM_Sec_2Mw_16_PRs_Rs_4} illustrates, for the same DBH
configuration, the scattering cross section at high frequency $2M_\infty\omega =16$.
In this case, with just $18$ Regge poles the glory and the orbiting oscillations are very
well reproduced. With $59$ Regge poles and without adding the background integral
the result obtained from the CAM approach is again indistinguishable from partial
wave expansion for $\theta \gtrsim 30^\circ $ on the plot.

Fig.\ \ref{fig:CAM_Sec_2Mw_16_Rs_5_PRs} shows the scattering cross section
for a DBH configuration with $R_s> 3M_\infty$ at $2M_\infty\omega = 16$.
Here, we have the shell configuration with $M_\infty=1.5 M_{\text{\tiny BH}}$ and
$R_s=5M_{\text{\tiny BH}}$.  In this case, to construct the result from CAM approach,
we  sum over three different branches and with $17$ Regge poles the glory as well
as the orbiting oscillations are captured. By including $48$ Regge poles, but no
background integral, the agreement is excellent with the partial wave expansion
result for  $\theta \gtrsim 30^\circ $.

Through Figs.~\ref{fig:CAM_Sec_2Mw_6_PRs_Rs_4},
\ref{fig:CAM_Sec_2Mw_16_PRs_Rs_4} and \ref{fig:CAM_Sec_2Mw_16_Rs_5_PRs},
we have shown that the scattering cross section can be described and understood in
terms of the Regge Poles, i.e., in terms of various contributions from different types
of resonances. In particular, we have identified different branches associated with
the properties of the light-rings and the shell position and/or its matter content.
To conclude our description of the differential scattering cross section, we return
to the discussion of critical effects associated with geodesic motion described in
Sec.\ \ref{sec:geodesics}. We can indeed understand qualitatively the observed
scattering cross sections from the various critical effect. As in the case of the
isolated black hole, the orbiting effect and the glory (associated with a single
light-ring) will result in oscillations in the scattering cross section as well as an
increase for $\theta \sim 180^\circ$.
In DBH spacetimes, we have shown that we may expect modulations due to the
rainbow effect and to grazing (creeping modes) or secondary orbiting. The
rainbow effect will lead to a local amplification of the scattering cross section
around the rainbow angle. Grazing, i.e., the edge rays (creeping modes)
or second orbiting will lead to further oscillations and global modulations of the
scattering cross section. This is illustrated in Fig.\ \ref{fig:comparison_critical}.
Note the qualitative difference that for configurations where the shell is outside
the outer light-ring, the scattering cross is mainly modulated around the rainbow
angle while for the case where the shell lies between the two light-rings the
scattering cross sections exhibits noticeable modulation at all angles.
Isolating and identifying quantitatively the different contribution from each
critical effect would require asymptotic formulae from the RPs and their
residues which are beyond the scope of this work.

\section{Conclusion and discussion}\label{sec:conclusion}

In this paper we have investigated the scattering of planar waves incident on
a dirty black hole spacetime, that is, a black hole surrounded by a thin shell of matter.
We have considered cases where the shell contributes substantially to the scattering,
thereby complementing the analyses presented in \cite{Macedo:2015ikq,Leite:2019uql}.
We have focused our attention on the spectrum of resonances of DBHs and calculated
the Regge pole spectrum of different DBH configurations.
This was done by extending the continued fraction method originally introduced by
Leaver, and following the approach of Ould El Hadj et al.\ \cite{OuldElHadj:2019kji}.

We have identified two key properties of the RP spectrum of DBHs:
i) the existence of two branches and ii) the possible existence of a third branch
with a nearly constant imaginary part.

The first two branches are associated with the inner light-ring  and with either the outer light-ring or the shell depending on the DBH configuration.
It appears that the two branches emerge from the original isolated black hole spectrum.
The separation from the original branch to the two distinct ones appears first for large overtone numbers while the two branches may overlap for low overtone. The separation is also more prominent for high frequencies and for DBH configuration where the shell contributes significantly. The existence of the two branches results in modulation of the scattering cross section. Such modulations were not seen in previous studies due to the specific DBH configuration considered by the authors.

In addition to these two branches, we have identified a third branch corresponding to broad resonances trapped between the inner light-ring and the thin shell. This interpretation was supported by a WKB analysis of the RP spectrum and is reminiscent of the resonances obtained for compact objects~\cite{Zhang:2011pq,OuldElHadj:2019kji,Volkel_2019}.

Building on the identification of the RP spectrum, we have constructed the complex angular momentum representation of the scattering. We have shown that the RP spectrum and their associated residues accurately describes the scattering cross section, including the modulations and deviations from the case of an isolated black hole. The accurate reconstruction of the scattering cross section also confirms that the resonances have been correctly identified and further validates our results.

Our study fits into the general goal of modelling and understanding black holes within their environments.
We have considered here a toy model of a black hole surrounded by a static and spherically symmetric thin shell of matter. In order to account for even more realistic scenarios, one would need to account for asymmetric distributions of matter and take into account the dynamical evolution of the environment.
The thin shell model provides a practical toy model for the latter as the dynamics of the shell can be found exactly. It would be an interesting extension of our work to investigate the resonances of dynamical dirty black holes. Indeed the literature of resonances of time dependent black hole is scarce with only numerical results which still lack a complete theoretical description~\cite{Abdalla:2006vb,Chirenti:2010iu}. We note the interesting recent developments in this direction~\cite{Bamber:2021knr,Lin:2021fth}.

Another interesting avenue to explore would be the impact of the environment
and its dynamics on the resonance spectrum is to mimic gravitational effects
using condensed matter platforms. This field of research, known as analogue
gravity~\cite{Unruh:1980cg}, has allowed us to observe and better understand
several effects predicted from field theory in curved spacetimes. Recently, analogue
simulators have investigated experimentally the ringdown phase of analogue black
holes~\cite{Torres:2020tzs}, and more experiments are being developed within the
Quantum Simulators for Fundamental Physics research programme
(https://www.qsimfp.org/). In analogue experiments, environmental and dynamical
effects will inherently be present and accurate modelling of their impact will be
necessary. These simulators will provide valuable insights to build the necessary
tools to describe dynamical dirty black holes.

\section*{Acknowledgments}

For the purpose of open access, the authors have applied a Creative Commons
Attribution (CC BY) licence to any Author Accepted Manuscript version arising.
There is no additional data associated with this article.
This work was supported in part by the STFC Quantum Technology Grants
ST/T005858/1 (RG \& TT), STFC Consolidated Grant ST/P000371/1 (RG).
SH is supported by King's College London through a KCSC Scholarship.
RG also acknowledges support from the Perimeter Institute.
Research at Perimeter Institute is supported by the Government of Canada
through the Department of Innovation, Science and Economic Development
Canada and by the Province of Ontario through the Ministry of Research,
Innovation and Science.

\bibliography{DBHCAMbibl}

\end{document}